\documentclass[10pt,twocolumn,oneside]{article}   	
\usepackage[colon]{natbib}

\usepackage[top=2cm, bottom=2cm, left=1.75cm, right=2cm]{geometry}
\usepackage{geometry}                		
\usepackage{graphicx}
\usepackage{dblfloatfix}
\usepackage[font=small,labelfont=bf]{caption}
\usepackage{abstract}
\usepackage{graphicx}				
\usepackage{amssymb}

\title{Membrane Trafficking of Integral Cell Junction Proteins and its Functional Consequences\footnote{Has not been submitted for publication}}

\author{Arie Horowitz\footnote{Correspondence: Faculty of Medicine, 22 Boulevard Gambetta, 76183 Rouen Cedex, Normandy, France; arie.horowitz@inserm.fr}\\{\textit{Faculty of Medicine, Normandy University, France}}}

\date{} 

\bibpunct{[}{]}{,}{n}{}{}

\begin{document}
\twocolumn[
\maketitle

\renewcommand{\abstractname}{\vspace{-\baselineskip}}

\vspace{-1.2cm}

\begin{onecolabstract}

\small{

Though membrane trafficking of cell junction proteins has been studied extensively for more than two decades, the accumulated knowledge remains fragmentary. The goal of this review is to synthesize published studies on the membrane trafficking of the five major junction transmembrane proteins: claudins, occludin, and junction adhesion molecules (JAMs) in tight junctions; cadherins and nectins in adherens junctions; to identify underlying common mechanisms; to highlight their functional consequences on barrier function; and to identify knowledge gaps. Clathrin-mediated endocytosis appears to be the main, but not exclusive, mode of internalization. Caveolin-mediated endocytosis and macropinocytosis are employed less frequently. PDZ-domain binding is the predominant mode of interaction between junction protein cytoplasmic tails and scaffold proteins. It is shared by claudins, the largest family of junction integral proteins, by junction adhesion molecules A, B, and C, and by the three nectins. All eight proteins are destined to either recycling via Rab4/Rab11 GTPases or to degradation. The sorting mechanisms that underlie the specificity of their endocytic pathways and determine their fates are not fully known. New data is presented to introduce an emerging role of junction-associated scaffold proteins in claudin membrane trafficking. 
}

\end{onecolabstract}

]
\saythanks

\begin{center}
\textbf{1. INTRODUCTION\footnote{Nomenclature: names of the proteins mentioned in the review follow current nomenclature rather than names used in the reviewed studies to increase clarity and uniformity (e.g., catenin p120 is named $\delta$-catenin).}}
\end{center}

Despite being no more than a single-cell thick, the endothelial and epithelial cell layers that form the luminal surface of blood and lymph vessels and of numerous tubular organs (e.g., the tracheal, digestive, and ureteric systems), respectively, are the only barriers that prevent breaching of the walls of each system and, concomitantly, its dysfunction. The cytoplasmic faces of adherens \citep{RN10}  and tight \citep{RN11}  junctions that maintain monolayer integrity harbor protein complexes which provide structural support and continuity from the junction transmembrane proteins to the cytoskeleton. These intricate molecular assemblies recycle constantly even in quiescent cells \citep{RN1,RN2,RN9} and undergo extensive remodeling in response to agonists such as vascular endothelial growth factor (VEGF), transforming growth factor (TGF)-$\beta$, sphingosine-1-phosphate (S1P), and platelet-derived growth factor (PDGF) \citep{RN3,RN6,RN7,RN520}. The molecular mechanisms that determine their membrane trafficking from and to the cell junctions are not fully known.

Survey of studies reported mostly during the last two decades reveal that despite the attention given to the subject of intercellular junction remodeling, the grasp of this process is fragmentary. Conceptual progress on cell junction dynamics is hampered by their structural and functional complexities. While there are only two or three species of junction transmembrane proteins in adherens \citep{RN23} and in tight junctions \citep{RN11}, respectively, some of them consist of protein families of multiple members. Moreover, they bind cytoplasmic protein complexes of varying compositions and sizes that mediate numerous signaling pathways and which interface with the cytoskeleton. A further layer of complexity is conferred by the diversity of membrane trafficking routes and endosomes encountered by different junction transmembrane proteins. 

Much of the data on cell junction dynamics reported to date was derived from epithelial cells and the organs they populate. Though it cannot be assumed by default that all the molecular mechanisms that control cell junction dynamics in epithelial cells are identical to the analogous mechanisms in endothelial cells (ECs), their observed similarities suggest that they are shared to a large extent. In a few cases, the review considers data obtained from fibroblasts used to express epithelial or EC junction proteins free of the confounding effects of the endogenous proteins. The review does not address desmosomes, which form in epithelial cells but not in ECs, or gap junctions, which have been studied primarily as electrophysiological interfaces. 

Intercellular junction proteins redistribute from the plasma membrane (PM) to the cytoplasm constitutively or in response to agonists such as VEGF in ECs \citep{RN4}, or epithelial growth factor (EGF) in epithelial cells \citep{RN163}. Once internalized, they have one of two possible fates: recycling back to the PM \citep{RN13,RN90}, or proteolysis in lysosomes or proteosomes \citep{RN509,RN110}. Membrane trafficking of the integral cell junction proteins confers the plasticity that cell junctions require to remodel in response to physiological and pathological stimuli.

For reasons that are related in part to the multiple functions of the catenins that bind their cytoplasmic domains \citep{RN477}, the membrane trafficking of cadherins is better known than that of all other intercellular junction transmembrane proteins, whereas that of nectins, the second adherens junction transmembrane protein species, is known the least.

The objective of this review is not to cover exhaustively all studies that have a bearing on membrane trafficking of junction proteins, but, to the extent possible, extract patterns and derive organizational principles. Because of the volume of the data and to maintain focus, the review addresses only the five major integral proteins of tight and adherens cell junctions, i.e., claudins, occludin, and JAMs, and cadherins and nectins, respectively. The cell systems, agonists, and experimental methods are specified in order to facilitate evaluation of the relevance and validity of the conclusions of the reviewed studies. 

The accompanying schemes in figures 1-5 represent syntheses of the reviewed studies on each of the five tight and adherens junction transmembrane proteins. They are intended to provide an ‘at a glance’ overall view of these pathways while maintaining the connection with the text by associating each component with one or more relevant studies. Figure 6 presents new data which indicates that the large junction-associated scaffold protein multiple PDZ domain protein (MPDZ) is involved in claudin membrane trafficking.

\begin{center}
\textbf{2. MEMBRANE TRAFFICKING OF TIGHT JUNCTION INTEGRAL PROTEINS}
\\[\baselineskip]
\textbf{2.1. CLAUDINS}
\end{center}

Encoded by 27 known human genes, claudins are the second largest family of intercellular junction transmembrane proteins \citep{RN47,RN46} after the cadherins. Claudins are a major tight junction structural component. Their variety may reflect tissue-specific expression of their genes, and/or differences in the extent of sealing they confer on tight junctions. Claudins are 210-305 amino-acid-long tetraspan proteins with cytoplasmic amino- and carboy-termini (Fig. 1). The majority of claudins harbor a post-synaptic density-95/Discs large/Zonula occludens (PDZ)-binding motif at their carboxy termini \citep{RN47}, implying that their recruitment to tight junctions or to trafficking pathways is mediated by proteins that harbor PDZ domains. Whereas most members of the claudin family genes are expressed in epithelial cells, their subsets vary among host organs. A smaller number of claudin genes are expressed in ECs. In mouse brain capillaries, the expression levels of Cldn5 and Cldn11 are predominant, but multiple sources indicate that Cldn1, 3, 10, 12 and 20 are also expressed at significant levels in these cells \citep{RN49}. Claudins seal the tight junctions by forming elongated peripheral strands on the cell’s surface \citep{RN103,RN508}. In endothelial \citep{RN254,RN338}, or epithelial \citep{RN157,RN309} cells, claudin removal by endocytosis from the tight junctions invariably increased monolayer permeability. Claudin strands on abutting cells can form heterophilic interactions, though only a small number of combinations had been reported \citep{RN80}. Claudin-1, a prototypical member of the family, underwent constitutive recycling in several epithelial cell types \citep{RN1} . Further observations recounted bellow suggested this behavior is shared by numerous claudins in epithelial and endothelial cells.

\begin{figure*}[h]
\center
\includegraphics[scale=0.6]{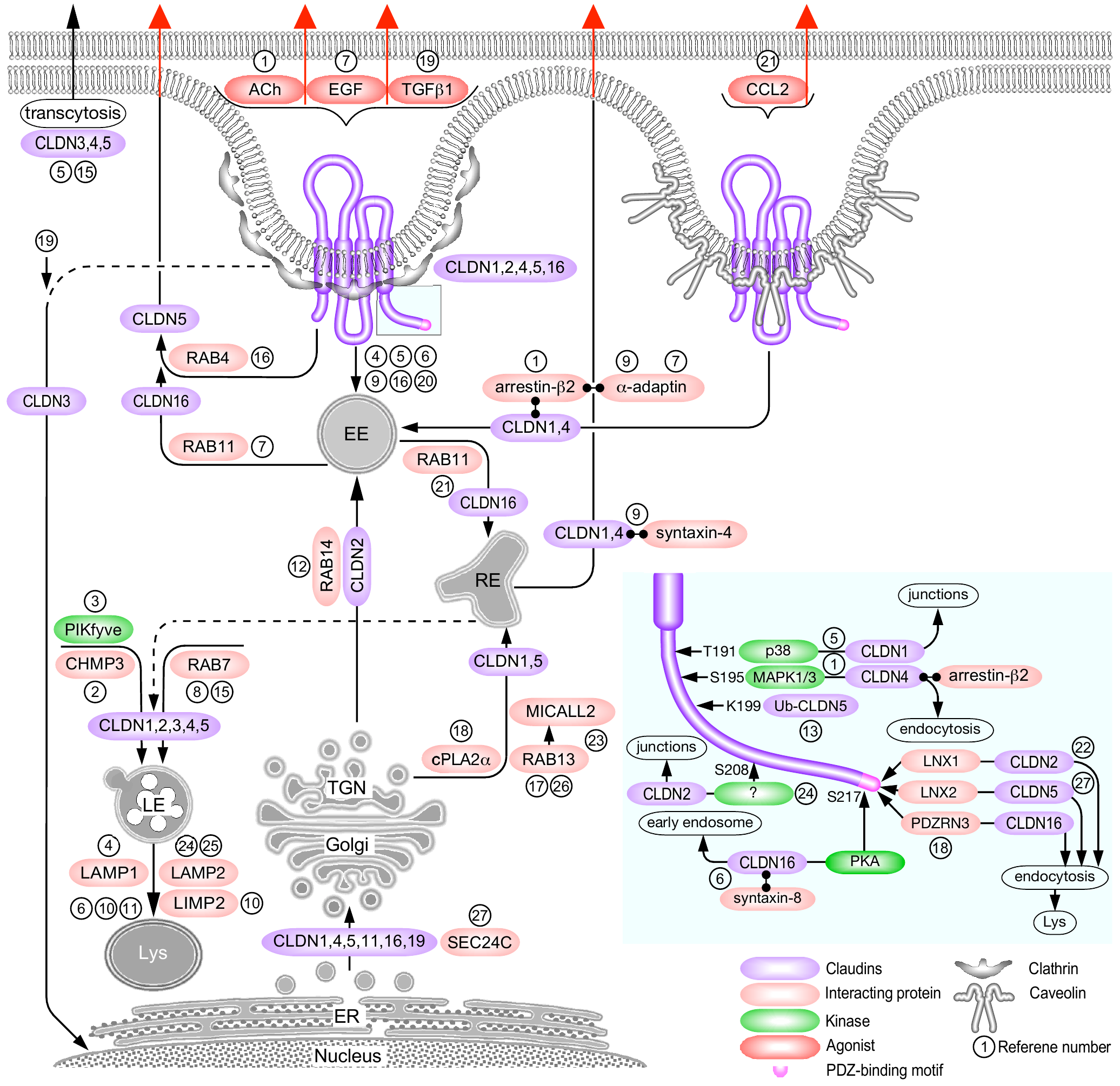}
\vspace{-0.3cm}

\caption{\textbf{Claudin membrane trafficking pathways.} Relevant studies are amalgamated rand coupled to the illustrated events by the circled numbers. Claudins were internalized via either clathrin caveolin or mediated endocytosis, or via macropinocytosis. It was partially retrieved to the PM via Rab4, Rab11 and Rab13-mediated recycling or diverted to lysosomal degradation. The claudin isoforms addressed in the reviewed studies are indicated. The inset presents phosphorylation sites that regulate claudin membrane trafficking, its carboxy-terminus PDZ-binding motif, and the ubiquitin ligases that bind it. EGF and CCL2-induced claudin endocytosis increased tight junction permeability (red arrows). The two adjoining PMs in this figure and in figures 2-5 represent two abutting cells. Dashed lines indicate speculative features. EE, early endosome; LE, late endosome; Lys, lysosome; RE, recycling endosome. Numbers correspond to the following references: (1) \citep{RN234}, (2) (\citep{RN1}, (3) \citep{RN38}, (4) \citep{RN256}, (5) \citep{RN233}, (6) \citep{RN268}, (7) \citep{RN243}, (8) \citep{RN35}, (9) \citep{RN14}, (10) \citep{RN51}, (11) \citep{RN281}, (12) \citep{RN157}, (13) \citep{RN52}, (14) \citep{RN296}, (15) \citep{RN54}, (16) \citep{RN235}, (17) \citep{RN70}, (18) \citep{RN44}, (19) \citep{RN478}, (20) \citep{RN254}, (21) \citep{RN53}, (22) \citep{RN60}, (23) \citep{RN99}, (24) \citep{RN37}, (25) \citep{RN28}, (26) \citep{RN102}, (27)  \citep{RN297}.}

\vspace{-0.3cm}

\end{figure*}

\textbf{2.1.1. Endocytosis}

The predominant endocytic pathway of the claudins is clathrin-mediated. Calcium depletion of human colon carcinoma epithelial cells resulted in claudin-1 and claudin-4 collocation with the clathrin heavy chain and with the clathrin adaptor protein $\alpha$-adaptin (AP2A1) \citep{RN367,RN14}. Their removal from the cell junctions was blocked by cytosolic acidification, sucrose-induced hypertonic stress, and phenylarsine oxide, all of which block clathrin-mediated endocytosis \citep{RN376,RN247,RN237}. The endocytosed claudins were recruited to early endosomes \citep{RN174}. Because claudins 1 and 4 collocated in that compartment with syntaxin-4, a \textit{t}-SNARE that mediates docking of transport vesicles to the PM \citep{RN242}, it appears they were readied for reincorporation in the PM once intercellular junctions were restored \citep{RN14}. In contrast to the effect of hypertonic stress, the induction of hypotonic stress by halving the osmolarity of the medium induced endocytosis of claudin-1 and -2 in Madin-Darby canine kidney (MDCK) cells. Its blockage by pharmacological inhibition of dynamin, a GTPase required for the scission of clathrin-coated vesicles \citep{RN365}, or of clathrin polymerization, indicated that claudin-1 and -2 underwent clathrin-mediated endocytosis \citep{RN256}. 

The homologous carboxy-terminus PDZ-binding motif shared by 20 members of the claudin family underlies the similarity of their membrane trafficking pathways. A missense mutation that replaced threonine in the -2 position of the motif to an arginine (T233R) was linked to familial hypomagnesaemia with hypercalciuria and nephrocalcinosis (FHHNC) \citep{RN236}, an inherited kidney disorder. Unlike native claudin-16, the mutant did not bind zona occludens (ZO)-1 when expressed in MDCK cells. When triggered to undergo constitutive endocytosis by temperature elevation from 4$\circ$C to 37$\circ$C, claudin-16 appeared in lysosomes instead of intercellular junctions. Though junction recruitment appeared to require the PDZ-binding motif, endocytosis evidently occurred in its absence because a claudin-16 with a missense L203X mutation, which truncates most of its intercellular carboxy-terminus domain including the PDZ-binding motif, was retained at the cell junctions when constitutive clathrin-mediated endocytosis was inhibited \citep{RN231,RN236}. However, the abundance of the L203X mutant at the cell junctions was substantially lower than the wild-type (WT) variant and, unlike the latter, was present throughout the cytoplasm (ibid.).

Clathrin-mediated endocytosis of claudins was induced by several physiological and pharmacological agonists. EGF induced claudin-2 binding to the clathrin heavy chain and $\alpha$-adaptin in MDCK epithelial cells \citep{RN243}. The endocytosed claudin-2 traversed early endosomes and collocated with lysosome-associated membrane protein (LAMP)-1 \citep{RN278}, but not with the Golgi apparatus, suggesting it was destined to lysosomal degradation rather than recycling. Cevimeline, a specific agonist of the M1 and M3 muscarinic acetylcholine (Ach) receptors \citep{RN511}, as well as carbachol, a non-selective Ach receptor agonist analog, induced claudin-4 binding to clathrin and to the endocytic sorting protein arrestin-$\beta$2 in immortalized rat salivary gland epithelial cells \citep{RN234}. This effect was specific to claudins, as the interaction of occludin and epithelial (E)-cadherin with arrestin-$\beta$2 was not altered by carbachol. The latter finding is in agreement with the M3 receptor specificity to arrestin-$\beta$2 \citep{RN249}. The clathrin dependence of claudin-4 endocytosis was established by showing that it was blocked by pharmacological inhibition of dynamin \citep{RN248}, by siRNA-mediated knockdown of the clathrin heavy chain, and by hypertonic sucrose medium. Endocytosed claudin-4 was ubiquintinated and underwent proteasomal degradation. Dissociation of \textit{trans}-binding claudins located on adjoining Caco-2 and human embryonic kidney (HEK)-293 epithelial cells by peptidomimetics of the first extracellular loop of claudin-1 induced clathrin-mediated endocytosis of endogenous claudin-1 and -5 \citep{RN254}. Because the endocytosis was a direct result of the release of the mechanical coupling between claudins rather than the downstream effect of specific agonists or cell culture conditions, this result is arguably the most generic indication that the default internalization mechanism of claudins is clathrin-mediated endocytosis. Uniquely among all the junction transmembrane proteins discussed in this review, claudin-3 redistributed from the tight junctions of primary human bronchial epithelial cells to their nuclei in response to TGF$\beta$1 downstream of TGF receptor 1 \citep{RN478}. The endocytic pathway and the nuclear entry mechanism of claudin-3 were not specified. Putative nuclear recognition signals had been identified in a majority of the claudins, and some of them had been observed in nuclei, but only in the context of cancer \citep{RN479}.

Aside from endocytosis into the parental cell, claudins underwent transcytosis into adjoining cells. During wound closure of monolayers of Eph4 or MDCK cells expressing GFP or FLAG (DYKDDDDK), respectively, fused to the amino-terminus of claudin-3 \citep{RN233}, cells of either type harbored cytoplasmic vesicles containing both endogenous claudin-3 as well as GFP or FLAG-fused claudin-3. This indicated that there was no preference between parental and transfected cells or between WT and recombinant claudins in the direction of transcytosis. The frequency of transcytosis was higher among dispersing subconfluent cells, suggesting that transcytosis resulted from mechanical tension along the intercellular junctions produced by cells pulling apart from each other. While the claudins were pinched off from the PM into tubular structures together with JAMA, occludin, and ZO1, they were sorted subsequently into claudin-only vesicles.

Constitutive transcytosis of claudin-4 or -5 overexpressed in MDCK cells, each fused at its amino-terminus to a different species of fluorescent protein, internalized simultaneously all the transmembrane proteins present on the transcytosed PM region . The endocytic vesicles contained both species of claudins as well as endogenous claudin-1, -2, or -7, or occludin \citep{RN233}. Single amino-acid substitutions F147A and Q156E in the second extracellular loop of claudin-5, which impair homophilic \textit{trans}-binding \citep{RN258}, reduced transcytosis frequency. Remarkably, transcytosis was inhibited by either chlorpromazine, an inhibitor of adaptor protein (AP)-2 complex binding to clathrin \citep{RN259}, or by filipin, which interferes with caveolae assembly by removal of cholesterol from the PM \citep{RN260}. Paradoxically, the dynamin-specific pharmacological inhibitor dynasore \citep{RN248} appeared to have no effect on transcytosis, in conflict with the effective chlorpromazine-induced inhibition. Collocation with the microtubule-associated protein 1A/1B-light chain 3 (MAP1LC3B) and with autophagy related protein 16L (ATG16L) indicated the transcytosed claudins were targeted to autophagosomes. Accumulation in the cytoplasm upon administration of the lysosome inhibitor chloroquine \citep{RN262} suggested that they were destined to lysosomal degradation, similar to the observation of Matsuda et al. \citep{RN54}.

While there is abundant evidence for the dependence of monolayer barrier function on claudin phosphorylation \citep{RN263,RN264,RN265,RN266,RN267}, the specific effect of phosphorylation on membrane trafficking is less documented. In quiescent MDCK cells, claudin-16 was phosphorylated constitutively \citep{RN268} by cyclic adenosine monophosphate (cAMP)-dependent protein kinase (PK)-A, as the phosphorylation was blocked by several PKA and adenylyl cyclase inhibitors and restored by cAMP. Single site mutations of potentially phosphorylatable serines pinpointed the PKA-phosphorylation to S217 in the claudin-16 carboxy-terminal cytoplasmic domain. Dephosphorylated claudin-16 did not associate with ZO1 and was located predominantly in the cytoplasm or in lysosomes. Claudin-16 phosphorylated on S217 was recruited to the cell junctions and bound the soluble N-ethylmaleimide-sensitive factor attachment protein receptor (SNARE) syntaxin-8 \citep{RN290}, which mediates endocytosis of cell membrane proteins and fusion of endocytic vesicles to early endosomes \citep{RN291}. The phosphorylation of claudins-3 and -5 was also attributed to PKA \citep{RN264,RN269}. S208 in the cytoplasmic domain of claudin-2 was phosphorylated in quiescent MDCK cells maintained in serum-supplemented medium \citep{RN37}. A phosphomimetic S208E mutant was detected at the cell edges, suggesting that S208 phosphorylation induced claudin-2 recruitment to cell junctions. Concordantly, the majority of the S208A dephosphomimetic mutant population was in the cytoplasm and collocated with the lysosomal marker protein LAMP2 \citep{RN278}. Mutations that reduced the recruitment of claudin-2 to the cell junctions, such as the removal of the carboxy-terminus PDZ-binding motif, were accompanied by a lower phosphorylation level, suggesting that the unidentified kinase that phosphorylated S208 was located on the cytoplasmic leaflet of the PM. 

Binding of claudin-4 to arrestin-$\beta$2 in carbachol-treated immortalized rat salivary gland epithelial cells was linked to the phosphorylation of S195 in the carboxy-terminus cytoplasmic domain of claudin-4 \citep{RN234}. The phosphorylation was attributed to MAPK1/3 because it was inhibited by PD98059, its pharmacological inhibitor \citep{RN251}, and by U0126, a pharmacological inhibitor of the phosphorylation and activation of MAPK1/3 by MAPKK1/2 \citep{RN252}. Claudin-4  phosphorylation on S195 was required for its binding to arrestin-$\beta$2, which, in turn, recruited clathrin. Hence, the phosphorylation was required for the clathrin-mediated endocytosis of claudin-4. Hypotonic stress-induced phosphorylation and endocytosis of claudin-1 and -2 in MDCK cells \citep{RN256} was attributed to MAPK p38 (MAPK14) because it was blocked by SB202190, its specific pharmacological inhibitor \citep{RN276}. Based on previous studies \citep{RN271,RN272,RN37}, the phosphorylated residues were assumed to be T191 in claudin-1 and S208 in claudin-2. This premise was supported by the resistance of the phosphomimetic claudin-1 and -2 mutants T191E and S208E, respectively, to their removal from the cell junctions by hypotonic stress. Conversely, the dephosphomimetic mutants T191A and S208A were segregated to the cytosol and collocated with the late endosome marker Rab7 and the lysosomal marker LAMP1.

Claudin-5 underwent caveolin/lipid raft-mediated endocytosis in mouse brain primary ECs in response to the monocyte chemoattractant \citep{RN449} chemokine C-C ligand (CCL)-2 \citep{RN53} based on the collocation of claudin-5 with cholera toxin, a marker of caveolin-mediated endocytosis and on the inhibition of claudin-5 endocytosis by cholesterol depletion. Collocation with either caveolin or a lipid raft marker was not shown.

\textbf{2.1.2. Recycling}

The recruitment of claudins from the Golgi apparatus to intercellular junctions required the activity of cytosolic phospholipase A2 (cPLA2)-$\alpha$, an arachidonic acid generating enzyme involved in the formation of cytoplasmic tubular membranes \citep{RN112}. In subconfluent human umbilical cord ECs (HUVECs) cPLA2$\alpha$ and claudin-5 were located in separate cytoplasmic punctae  \citep{RN44}. There was a reciprocal relationship between the cellular locations of cPLA2$\alpha$ and claudin-5: when the cells reached confluence, inactive cPLA2$\alpha$ was present in the Golgi apparatus \citep{RN113}, whereas claudin-5 was located at intercellular junctions. Depletion of cPLA2$\alpha$ by knockdown of its gene, \textit{PLA2G4A}, or by pyrrolidine-mediated pharmacological inhibition \citep{RN294}, was accompanied by removal of claudin-5 from the cell junctions and its accumulation in the Golgi apparatus, suggesting that cPLA2$\alpha$ activity was required for claudin-5 trafficking from the Golgi to the junctions.

Rab13 plays a major role in the recruitment of tight junction transmembrane proteins to cell junctions \citep{RN282}. The recycling of claudin-1 to the PM after recovery of MDCK cells from calcium depletion was slowed down by siRNA-mediated knockdown of Rab13, or by expression of the GTP hydrolysis-defective mutant Rab13Q67L \citep{RN102}. The recycling required Rab13 binding to its effector molecule interacting with CasL-like 2 (MICALL2) \citep{RN99}, a filamentous (f)-actin and $\alpha$-actinin-4 binding protein that drives f-actin crosslinking during intercellular junction assembly \citep{RN283,RN284}. The trafficking pattern of Rab13 suggested that claudin-1 translocated from the trans-Golgi network (TGN) to recycling endosomes, and from them to the PM \citep{RN70}.

Claudins employ several trafficking pathways and multiple Rab GTPases that are determined in part by the conditions the cells are subjected to. Claudin-5 endocytosed in mouse brain microvascular ECs in response to CCL2 \citep{RN340}, and collocated with Rab4 \citep{RN53}, a ‘fast’ trafficking GTPase \citep{RN280}. It did not collocate with the lysosomal marker LAMP2, indicating it was recycled to the PM rather than degraded. In agreement, the recycling inhibitor bafilomycin-A1 \citep{RN394} prevented the removal of claudin-5 from Rab4-containing vesicles. Claudin-16 recycling in quiescent MDCK cells was regulated by Rab11 \citep{RN35}, a ‘slow’ recycling GTPase \citep{RN286}. Inhibition of Rab11 activity by either a dominant-negative Rab11S25N mutant, or by primaquine, a pharmacological inhibitor of vesicle trafficking \citep{RN287} decreased claudin-16 presence at the cell junctions in favor of collocation with the early endosome antigen (EEA) 1 and the lysosome marker lysosome membrane protein (LIMP)-2 \citep{RN288}. The recycling of claudin-2 from the cytoplasm to the junctions of MDCK cells switched from incubation on ice to 37$\circ$C required the activity or Rab14 \citep{RN157}, a GTPase involved in trafficking from the TGN to early endosome \citep{RN289}. Knockdown of Rab14 by short hairpin (sh) RNA resulted in claudin-2 targeting to lysosomes. Rab7, a late endosome marker \citep{RN241}, mediated the trafficking of transcytosed claudin-3 and -4 to the lysosome in MDCK cells \citep{RN54}. 

Similar to around a third of all translated proteins in eukaryotic cells \citep{RN300}, the translocation of claudin-1 from the endoplasmic reticulum (ER) to the Golgi apparatus occurs via membrane trafficking in coat protein complex (COP)-2 vesicles. The binding of claudin-1 to the Sec24C subunit of COP2 \citep{RN301} was shared by claudins 4, 5, 11, 16, and 19 \citep{RN297}, all of which have a tyrosine and valine at their carboxy-termini that functions as an ER export signal. These residues were required but not sufficient for Sec24c binding, which probably involves the whole claudin cytoplasmic domain. Expectedly, knockdown of Sec24c reduced the abundance of claudin-1 at the cell surface.

The endosomal sorting complex required for transport (ESCRT), a multi-protein assemblage that mediates membrane budding and scission in several cellular contexts \citep{RN304} and the sorting of ubiquitinated cargo \citep{RN307} was required for the constitutive recycling of claudin-1 in quiescent confluent MDCK cells \citep{RN1}. Expression of truncated CHMP3, an ESCRT3 component (Raiborg and Stenmark, 2009), blocked membrane budding \citep{RN306}, resulting in the mutant’s accumulation in abnormally large vacuolar structures that contained both early and late endosomal markers. Dominant- negative CHMP3 collocated with claudin-1 and -2 and with ubiquitin. Because surface biotinylation indicated a reduction of cell surface claudin-1, it was concluded that dominant-negative CHMP3 impaired claudin-1 recycling. However, the ESCRT complex is thought to mediate primarily the budding of multivesicular body (MVB) intraluminal vesicules \citep{RN307}, the main source of lysosome-targeted cargo, rather than recycling tubules. Since the cell-surface biotin fraction was not separated from the rest of the cell lysate, it is conceivable that the abundance of recycled claudin-1 was overestimated. The interaction of CHMP3 with phosphatidylinositol (3,5) phosphate (PtdIns(3,5)P2), an endomembrane phospholipid, is essential for MVB genesis \citep{RN334}. PtdIns(3,5)P2) and its precursor, PtdIns 5-P, are synthesized by phosphatidylinositol 3-phosphate 5-kinase (PIKfyve) \citep{RN335}. Concordantly, perturbation of PIKfyve activity by the pharmacological inhibitor YM201636 demonstrated that constitutive recycling of claudin-1 in MDCK cells depends on PIKfyve \citep{RN38}, similar to its dependence on ESCRT \citep{RN302}. Approximately 35 percent of total cell surface claudin-1 was endocytosed and recycled in its majority back to the PM in untreated MDCK cells. In contrast, administration of YM201636 resulted in the accumulation of all the claudin-1 population in large cytoplasmic clusters. Whereas claudin-2 responded to YM201636 like claudin-1, the trafficking of claudin-4 was unaffected, suggesting that its rate of endocytosis is significantly lower than those of claudin-1 and -2. The inhibition of claudin-1 and -2 endocytosis prevented restoration of the barrier function of MDCK cell monolayers \citep{RN38}. Collectively, these results indicate that endocytic trafficking pathways differ not only among tight junction protein species but also within the claudin family.

Though most of the data on claudin membrane trafficking addressed individual claudin species, their dynamics are interdependent. This is not surprising given the large number of claudin species and the abundance of several species in the same cell. Claudin-4 trafficking in quiescent HEK-293 cells depended on claudin-8, but not vice-versa \citep{RN232}. Normally the two claudins traveled together in endocytic vesicles and bound the scaffold protein multi (M)PDZ \citep{RN292,RN293}. However, when claudin-8 was knocked down by siRNA, claudin-4 was sequestered to the ER and the Golgi apparatus. The intracellular dynamics and cell junction recruitment of claudins-2 and -4 in and MDCK cells differed from each other \citep{RN28}. In confluent cells, both claudins were located mainly at the cell junctions and to a lesser extent in cytoplasmic vesicles. Both newly synthesized claudins originated in the Golgi apparatus, but claudin-4 preceded claudin-2 at the cell junctions. Conjugation of fluorophores that emitted at either 549 or 505 nm to ‘old’ or to newly synthesized claudins, respectively, revealed that ‘old’ claudins were removed from the cell junctions to endocytic vesicles. Part of these vesicles were destined to lysosomal digestion, as they collocated with LAMP2. Surprisingly, removal of the carboxy-terminus PDZ-binding motif of claudin-4, which is required for binding to the PDZ domains of ZO1 and ZO2 \citep{RN298}, slowed the rate but not the steady-state abundance of claudin-2 at the cell junctions. Apparently, the PDZ-binding motif facilitates but is not categorically required for claudin recruitment to cell junctions. The truncated claudin-4 half-life was longer than that of WT claudin-4, likely because it was unable to bind the E3 uniquitin ligase ligand of numb-protein X (LNX)1. The removal of the PDZ-binding motif of claudin-2 probably did not abolish its binding to ZO1 or ZO2 because, unlike claudin-1 and 4, it harbors a tyrosine at position -6 that is required for the formation of a second ZO binding site \citep{RN299}. This tyrosine, which is present in other eight claudins, may account for the overabundance of claudin-2 over claudins 1, 3, 4, and 7 at the cell junctions of MDCK cells \citep{RN28}.

\textbf{2.1.3. Degradation}

Endocytosed claudins, either constitutively or in response to external stimuli, undergo either lysosomal digestion or proteosomal degradation. Claudin-1 and -2 endocytosed in MDCK cells subjected to hypotonic stress and treated by the lysosome inhibitor chloroquine accumulated in LAMP1-associated endosomes \citep{RN256}. The inhibition of this accumulation by SB202190 indicated it depended on p38-dependent trafficking, though the specific role of p38 in this process was not reported. Deprivation of oxygen and glucose from-immortalized bEND3 mouse brain ECs, was followed by claudin-5 lysosomal degradation rather than recycling \citep{RN51,RN281}. In quiescent Henrietta Lax (HeLa) cervical carcinoma cells, claudin-5 was polyubiquitinated on K199, K214, and K215 in the carboxy-terminus cytoplasmic domain, though ubiquitination of K199 was sufficient for proteolysis of more than half of the cellular claudin-5 population \citep{RN52}. Several claudins were ubiquitinated by LNX1, a protein that harbors 4 PDZ domains. The first PDZ domain of LNX1 bound the claudin-1 PDZ-binding motif in quiescent MDCK cells \citep{RN60}. Overexpression of EGFP-fused LNX1 resulted in the removal of claudins 1, 2, and 4 from the cell junctions. EGFP-LNX1collocated in part at the cell junctions with ZO1, occludin, and E-cadherin, whereas it collocated with claudin-2 in the cytoplasm, including in late endosomes and lysosomes. Apparently, LNX1 ubiquitinated only claudins at the cell junctions. Since ubiquitination serves as both an endocytic signal and for targeting to the interior of MVBs \citep{RN295}, LNX1 may have designated cell junction claudins for removal from the cell junctions by endocytosis, followed by lysosomal digestion. LNX2, which has the same domain structure as LNX1 and is close to fifty percent identical, bound the carboxy-terminus of claudin-1 \citep{RN297} and was likely, therefore, to have functioned as its E3 ubiquitin ligase. Rather than LNX1 or 2, claudin-16 was ubiquitinated by the two PDZ domain-containing RING finger protein (PDZRN)-3 \citep{RN296}. Similar to LNX1, PDZRN3 bound claudin-16 via its PDZ-binding motif, induced its endocytosis from the cell junctions, and designated it to lysosomal digestion, as indicated by its collocation with the late endosome marker Rab7.

\begin{center}
\textbf{2.2. OCCLUDIN}
\end{center}

Occludin is encoded by a single gene in humans. It is grouped into a 3-member proteins family named tight junction–associated MARVEL protein (TAMP), based on sharing a ~130-residue MAL and related proteins for vesicle trafficking and membrane link (MARVEL) domain \citep{RN382}. The MARVEL region spans the four transmembrane domains, corresponding to residues 60-269 of human occludin. Similar to claudins, occludin is a tetraspan protein, but its size is substantially larger (522 amino acids,), primarily due to a longer carboxy-terminus domain that consists of 257 amino acids in humans (Fig. 2). Its cytoplasmic domain binds the Src homology 3 (SH3) and the guanylate kinase (GuK) domains of ZO1 via its coiled-coil region \citep{RN76,RN77}. Occludin is recruited to the tight junctions by claudin \citep{RN48} and incorporated into the claudin strands in a punctate pattern \citep{RN312}. 

\begin{figure*}[h]
\center
\includegraphics[scale=0.6]{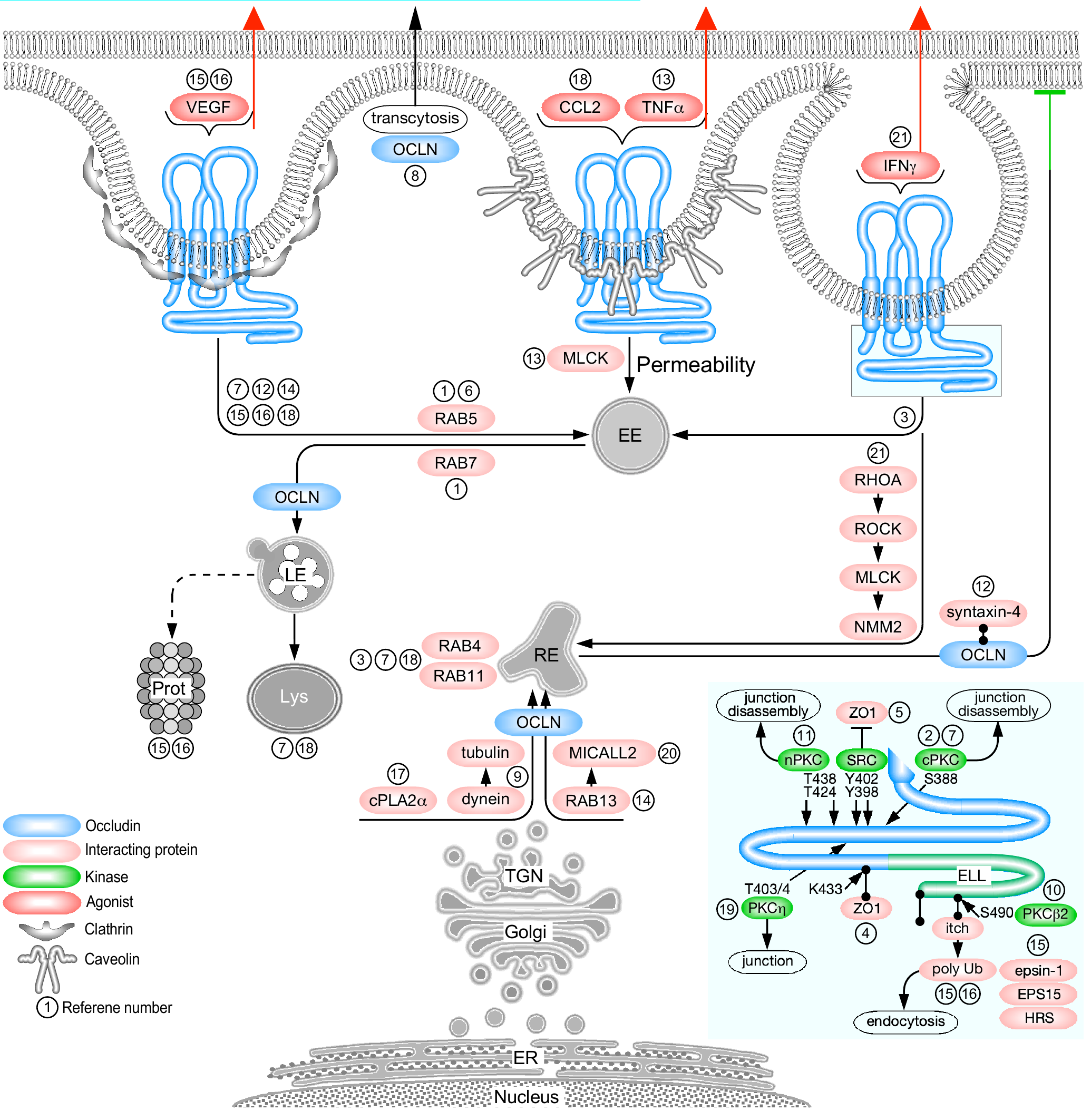}

\vspace{-0.0cm}
\caption{\textbf{Occludin membrane trafficking pathways.} Occludin was internalized via either clathrin or caveolin-mediated endocytosis, or via macropinocytosis. It was partially retrieved to the PM via Rab4, Rab11 and Rab13-mediated recycling or diverted to lysosomal or proteasomal degradation. The inset presents phosphorylation sites that regulate occludin membrane trafficking and binding sites of ZO1 and itch. VEGF, CCL2, TNF$\alpha$,, and IFN$\gamma$-induced occludin endocytosis increased tight junction permeability (red arrows). Rab13-mediated occludin recycling restored tight junction integrity (green T). Dashed line indicates speculative feature. EE, early endosome; LE, late endosome; Lys, lysosome; MLCK, myosin light chain kinase, NMM2, non-muscle myosin 2; Prot, proteasome; RE, recycling endosome. Numbers correspond to the following references: (1) \citep{RN456}, (2) \citep{RN321}, (3) \citep{RN100}, (4) \citep{RN310}, (5) \citep{RN68}, (6) \citep{RN454}, (7) \citep{RN2}, (8) \citep{RN233}, (9) \citep{RN452}, (10) \citep{RN317}, (11) (\citep{RN322}, (12) \citep{RN14}, (13) \citep{RN57}, (14) \citep{RN13}, (15) \citep{RN7}, (16) \citep{RN63}, (17) \citep{RN44}, (18) \citep{RN53}, (19) \citep{RN69}, 20) \citep{RN99}, (21) \citep{RN101}.}

\vspace{-0.5cm}

\end{figure*}

\textbf{2.2.1. Endocytosis}

There is extensive similarity between the endocytic and recycling pathways of claudins and occludin. Like claudins, occludin underwent continuous constitutive endocytosis in malignant MTD-1A mouse mammary epithelial cells \citep{RN13}. In a monolayer of MDCK cells that underwent wounding, occludin underwent clathrin-mediated endocytosis with a half-time of 15 min \citep{RN2}. A previous study from the same group established that occludin endocytosis was mediated by clathrin \citep{RN454}. Similar to claudin-1 and -4, occludin collocated with the clathrin heavy chain and with the clathrin adaptor $\alpha$-adaptin in human colon carcinoma epithelial cells after calcium depletion \citep{RN14}. VEGF induced clathrin-dependent endocytosis and phosphorylation of S490 in the carboxy-terminus cytoplasmic domain of occludin in primary bovine retinal ECs \citep{RN7}. The phosphorylation induced binding of the E3 ligase Itch and prompted occludin poly-ubiquitination, ubiquitination, though the underlying mechanism remained speculative because of the large distance between S490 and the polyp-proline motif in occludin’s amino-terminus to which the WW domain of Itch binds \citep{RN55}.  Ubiquitinated occludin bound epsin-1 epidermal growth factor receptor pathway substrate (EPS)-15, and hepatocyte growth factor-regulated tyrosine kinase substrate (HRS). Both proteins harbor ubiquitin-interacting motifs and facilitate endocytosis of ubiquitinated cell-surface proteins \citep{RN316}. Endocytosed occludin collocated with each of these proteins in cytoplasmic punctae. Rab5, an early endosome marker \citep{RN455}, also collocated with occludin in cytoplasmic punctae in a monolayer of wounded MDCK cells \citep{RN454}.

TNF$\alpha$, the prototypic member of the tumor necrosis factor ligand superfamily, induced caveolin-1-mediated occludin endocytosis in mouse jejunal epithelial cells downstream of myosin light chain kinase (MLCK) activation \citep{RN57} and in Caco-2 and T84 human colon metastatic epithelial cells \citep{RN310}. The endocytosis required the carboxy-terminal 107 amino acids of the cytoplasmic tail of occludin, TNF$\alpha$ induced caveolin-mediated endocytosis of occludin in Caco-2 and T84 human colon metastatic epithelial cells. The endocytosis required the carboxy-terminal 107 amino acids of the cytoplasmic tail of occludin, a region shared with and named after RNA polymerase II elongation factor ELL, a region shared with and named after RNA polymerase II elongation factor ELL \citep{RN312}. The ELL region afforded binding to ZO1 and occludin dimerization \citep{RN76}. In its absence, truncated occludin remained at the lateral cell membrane despite TNF$\alpha$ treatment, and, inversely, its expression had a dominant-negative effect on the endocytosis of endogenous occludin \citep{RN310}. The ELL positively charged residues, in particular K433, were essential for ZO1 binding to occludin \citep{RN310}.

All three major tight junction transmembrane proteins, claudin, occludin, and JAM, underwent macropinocytosis in T84 cells treated by interferon (IFN)-$\gamma$ \citep{RN101}, a cytokine that disrupts the integrity of epithelial cell monolayers \citep{RN339}. The internalized tight junction transmembrane proteins were recruited collectively to subapical actin-coated vacuoles at a relatively slow rate, 38 hours after IFN$\gamma$ treatment. IFN$\gamma$ initiated the formation of vacuoles by activating non-muscle myosin-2 via RhoA, Rho-associated kinase, and MLCK. The vacuoles were identified as recycling endosomes by collocation with Rab4 and Rab11 \citep{RN100}. Like claudin-5, occludin collocated with cholera toxin, indicating it underwent caveolin/lipid raft-mediated endocytosis in CCL2-treated mouse brain primary ECs \citep{RN53}. IFN$\gamma$-induced endocytosis of occludin reduced the trans-epithelial resistance of monolayers of Caco-2 cells \citep{RN56}.

Similar to claudins, occludin is phosphorylated at multiple sites in the carboxy-terminus cytoplasmic domain. The phosphorylation's effects on occludin endocytosis were site-specific. The most predominant occludin kinases were either ‘conventional’ (c) Ca\textsuperscript{2+} and diacylglycerol (DAG)-dependent, or ‘atypical’ (a) Ca\textsuperscript{2+} and DAG-independent PKCs. Because DAG induces tight junction assembly \citep{RN319}, the DAG analogs phorbol 12-myristate 13-acetate and 1,2-dioctanoylglycerol were administered to MDCK cells to activate cPKCs \citep{RN321}. Under low calcium conditions, the treatment resulted in occludin’s phosphorylation on S388 (detected by mass spectroscopy) and induced its recruitment to tight junctions. Conversely, treatment of Ca\textsuperscript{2+}-replenished MDCK cells with the GF-109203X PKC-specific pharmacological inhibitor \citep{RN320} resulted in occludin’s dephosphorylation. The phosphorylation detected by mass spectroscopy on S388 was attributed to a cPKC because a mixture of PKC$\alpha$, $\beta$I, $\beta$II, and $\gamma$ phosphorylated in vitro the recombinant carboxy-terminus cytoplasmic domain of occludin. In contrast, calcium replenishment induced phosphorylation of T403/404 and promoted the recruitment of occludin to MDCK cell junctions \citep{RN69}. Occludin mutated on T403/404 to dephosphorylation-mimicking alanines was removed from the cell junctions upon calcium replenishment, whereas phosphorylation-mimicking T403/404D mutants were recruited to the junctions. Motivated by its conspicuous expression in epithelial cells \citep{RN318}, immunoblotting of occludin-precipitated cell lysate confirmed that PKC$\eta$, a ‘novel’ diacylglycerol-dependent PKC, bound the carboxy-terminus cytoplasmic domain of occludin. Its pharmacological inhibition by a pseudo substrate peptide \citep{RN505} and by its knockdown resulted in the disruption of the tight junctions of MDCK cells. Intriguingly, a study from the same group concluded that calcium replenishment-induced phosphorylation of the nearby Y398 and Y402 of human occludin expressed in MDCK cells had an opposite effect on occludin: rather than recruiting occludin, it prevented its incorporation in cell junctions \citep{RN68}. This effect was attributed to the abolishment of occludin binding to ZO1. In vitro experiments suggested that Src was the likely kinase responsible for the phosphorylation.

The involvement of aPKCs in the maintenance of epidermal barrier function \citep{RN322} instigated investigation into the role of PKC$\zeta$ in the regulation of Caco-2 cell tight junctions \citep{RN512}. Treatment of quiescent cell monolayers by a myristoylated PKC$\zeta$ pseudosubstrate disrupted the tight junctions and slowed their assembly when the cells were subjected to a ‘calcium switch’, i.e., incubation in a calcium-chelating medium followed by calcium-enriched medium. PKC$\zeta$ bound a recombinant 150 residue-long recombinant carboxy-terminus cytoplasmic domain of occludin in vitro. Threonine scanning of the cytoplasmic domain attributed the phosphorylation to T424 and T438 in the carboxy-terminus cytoplasmic domain.

Calcium replenishment-induced phosphorylation of Tyr398 and Tyr402 of human occludin expressed in MDCK cells prevented its recruitment to cell junctions and its binding to ZO1 \citep{RN68}. In vitro experiments suggested that Src was the likely kinase responsible for the phosphorylation. In contrast to the effects of the above phosphorylations, calcium replenishment-induced phosphorylation of the nearby T403/404 sites (determined by mass spectroscopy) induced occludin recruitment to the intercellular junctions of MDCK cells \citep{RN69}. Occludin mutated on T403/404 to dephosphorylation-mimicking alanines was removed from the MDCK cell junctions upon calcium replenishment, whereas mutants harboring phosphorylation-mimicking replacements to aspartate were recruited to the junctions. Motivated by its conspicuous expression in epithelial cells \citep{RN318}, immunoblotting of occludin-precipitated cell lysate confirmed that PKC$\eta$, a ‘novel’ diacylglycerol-dependent PKC, bound the carboxy-terminus cytoplasmic domain of occludin. Its pharmacological inhibition by a pseudo substrate peptide \citep{RN505} and knockdown resulted in the disruption of the tight junctions of MDCK cells.

The effects of VEGF downstream signaling on intercellular junction proteins have been of obvious interest because the growth factor induces intercellular junction disassembly \citep{RN324}. Since a phosphorylation cascade typically ensues downstream of tyrosine kinase receptors, numerous studies probed the extent of the phosphorylation of intercellular junction transmembrane proteins. Several early studies detected VEGF-induced phosphorylation of occludin in primary bovine retinal ECs \citep{RN61} and determined that it existed in five to seven phosphorylated forms, suggesting the presence of multiple phosphorylation sites that were at least in part the substrates of PKC$\beta$2 \citep{RN317}. Subsequent mass spectroscopy analysis of VEGF-treated bovine retinal ECs detected phosphorylation on S490 \citep{RN62}. VEGF increased occludin abundance in the cytoplasm at the expense of its presence along the cell borders. PKC$\beta$1, an ATP-competitive PKC$\beta$-specific inhibitor, reversed the VEGF-induced redistribution of occludin, suggesting that it was triggered by S490 dephosphorylation \citep{RN63}.

\textbf{2.2.2. Recycling}

Like claudin, Rab13 played a major role in the constitutive recycling of occludin in quiescent malignant mouse mammary epithelial cells, as indicated by the stabilization of occludin at the cell junctions by expression of a dominant-negative Rab13Q67L mutant in MTD-1A cells. Both occludin and claudins 1 and 4 collocated with the \textit{t}-SNARE protein syntaxin-4 in T84 epithelial cells \citep{RN14} and required the Rab13 effector MICALL2 to recycle back to the PM \citep{RN99}. Deletion of the Rab13-binding domain of MICALL2 disrupted occludin recycling to the PM and prevented an increase in the trans-epithelial resistance, indicating that the permeability of the cellular monolayer remained high. Whereas Rab13-dependent trafficking bypassed the canonic recycling markers Rab4 and Rab11 \citep{RN13}, occludin collocated with Rab4 in mouse brain primary ECs treated by CCL2 \citep{RN53} and with both Rab4 and Rab11 in IFN$\gamma$-treated T84 cells \citep{RN100}. In quiescent serum-starved MDCK cells, a minority of the claudin population returned to the PM via recycling endosomes in a Rab11-dependent manner \citep{RN2}. Furthermore, cPLA2$\alpha$ regulated the trafficking of occludin between the Golgi apparatus and HUVEC junctions in the same manner as claudin-5: depletion or inhibition of cPLA2$\alpha$ was accompanied by the removal of occludin from the cell junctions and its accumulation in the Golgi apparatus \citep{RN44}. Possibly because of its role in the maintenance of the Golgi apparatus structure \citep{RN450}, partitioning defective protein 3 (Par-3) was required for the trafficking of occludin from the TGN to the junctions of TNF$\alpha$-treated Caco-2 cells. In its absence, occludin accumulated in the TGN \citep{RN451}. Occludin translocated from the TGN to the surface of quiescent Caco-2 or MDCK cells along microtubules in vesicles propelled by the minus-end-directed molecular motor dynein at an approximate velocity of 1.6 $\mu$m/s \citep{RN452}. The first 18 amino acids of the cytoplasmic domain were sufficient for targeting occludin to the cell surface. Among these, residues I279 and W281 were essential, possibly because they constituted a TGN export signal \citep{RN453}.

\textbf{2.2.3. Degradation}

Unlike claudins, ESCRT did not regulate the recycling and fate of occludin \citep{RN1}. Though there is heterogeneity in the half lives of claudins \citep{RN329}, several studies concurred that the half-life of occludin is shorter than those of claudins by as much as three-fold \citep{RN328,RN55}. Direct measurements of occludin and claudin-1 dynamics in quiescent MDCK cells by fluorescence recovery after photobleaching revealed an inverse relation between the two proteins: whereas 76 percent of the steady-state population of claudin-1 was attached to cell junctions, the size of the unattached cytoplasmic fraction of occludin under the same conditions was 71 percent \citep{RN9}.

The aforementioned ubiquitin ligase Itch bound the amino-terminus cytoplasmic domain of occludin in HEK-293 and LLC-PK pig kidney epithelial cells \citep{RN55}. Based on the effect of the proteasome-specific inhibitor MG132 \citep{RN331}, occludin was determined to undergo proteasomal degradation. The VEGF-induced phosphorylation of occludin on S490 in bovine endothelial retinal cells discussed above was required for Itch binding to occludin, for occludin ubiquitination, and, subsequently, for its proteasomal degradation \citep{RN7}. VEGF augmented an ongoing low-level constitutive proteasomal degradation of occludin. In serum-starved MDCK cells, close to half of the cellular occludin population underwent constitutive endocytosis in an approximate half-time of 15 min. The majority of the endocytosed occludin population was apparently degraded, as only 20 percent returned to the cell surface \citep{RN2}. Based on collocation with neuropeptide-Y, a lysosomal marker \citep{RN333}, and on the effect of bafilomycin-A1, a lysosomal inhibitor \citep{RN332}, occludin degradation was attributed to the lysosome.

\begin{center}
\textbf{2.3. JUNCTIONAL ADHESION MOLECULE (JAM)}
\end{center}

The best-known members of the JAM protein family are encoded by 3 genes, F11R, JAM2 and JAM3. JAM’s are 298-310 amino acid-long single pass proteins of the immunoglobulin family (Fig. 3). Their cytoplasmic domains range in length from 39 to 48 amino-acids and harbor a PDZ-binding motifs at their carboxy-termini that binds the scaffold protein ZO1 at intercellular junctions \citep{RN336}. JAM-A forms homophilic \textit{trans}-dimers \citep{RN337}, whereas JAM-B and JAM-C can form either homo- or heterophilic interactions \citep{RN81}. The topology of the JAMs as single pass transmembrane proteins contrasts with the four-pass topology of both claudins and occludin. A substantial part of the relatively limited number of studies, compared to the latter two proteins, was focused on virus entry (e.g., \citep{RN338}), a phenomenon that is deliberately not addressed in this review. Consequently, knowledge of the membrane trafficking of the JAM proteins under physiological conditions is relatively scarce, particularly of JAM-B and -C. Though multiple residues in the JAM cytoplasmic domain are phosphorylated by identified kinases \citep{RN467}, their phosphorylation has not been linked to membrane trafficking.

\begin{figure*}[h]
\center
\includegraphics[scale=0.60]{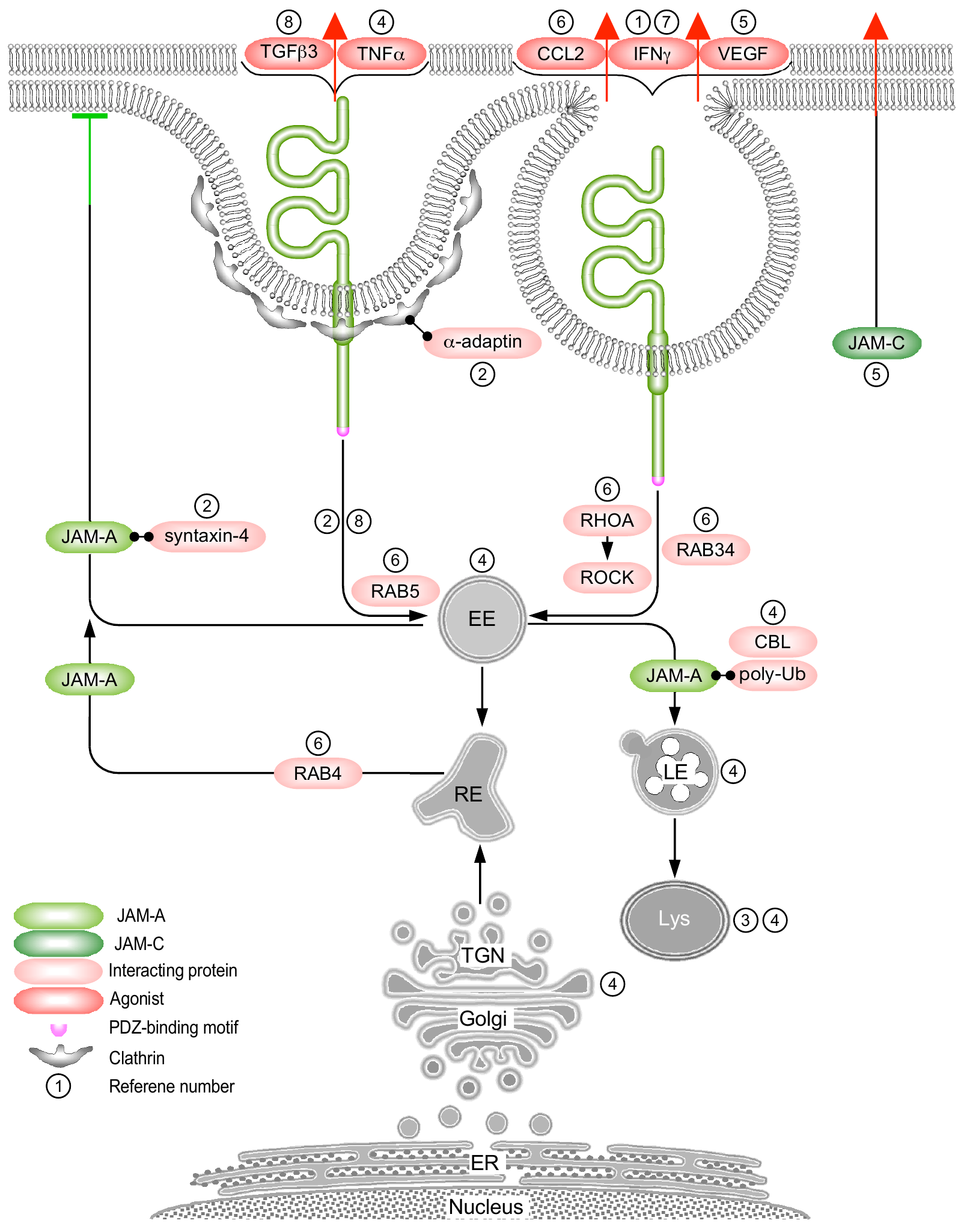}
\vspace{-0.0cm}
\caption{\textbf{JAM membrane trafficking pathways.} JAMs were internalized via clathrin-mediated endocytosis or via macropinocytosis. They were partially retrieved to the PM via Rab4-mediated recycling or diverted to lysosomal degradation. JAM removal from the PM increased tight junction permeability. TGF$\beta$3, TNF$\alpha$, CCL2, and IFN$\gamma$-induced JAM-A/C endocytosis increased tight junction permeability (red arrows). EE, early endosome; LE, late endosome; Lys, lysosome; RE, recycling endosome. Numbers correspond to the following references: (1) \citep{RN100}, (2) \citep{RN14}, (3) \citep{RN517}, (4) \citep{RN515}, (5) \citep{RN342}, (6) \citep{RN53}, (7) \citep{RN101}, (8) \citep{RN468}.}

\vspace{-0.5cm}

\end{figure*}

Several of the studies reviewed above analyzed multiple species of tight junction proteins, including the JAMs. Similar to claudins-1 and -4, JAM-A collocated with the clathrin heavy chain and $\alpha$-adaptin in T84 cells, indicating that they were all endocytosed via clathrin-coated vesicles \citep{RN14}. Subsequently, JAM-A was recruited to early endosomes, followed by segregation into a cytoplasmic subapical compartment where it collocated with syntaxin-4, possibly in preparation of reincorporation in the PM upon restoration of intercellular junctions. Collectively, these observations imply that JAM-A shares the endocytic pathway of the other prominent tight junction proteins in response to calcium chelation. The same conclusion can be drawn in regard to the administration of IFN$\gamma$ to T84 cells, whereby both occludin and JAM-A underwent macropinocytosis \citep{RN100,RN101}. The internalized proteins subsequently collocated with markers of early endosomes and with Rab4 and -11, markers of ‘fast’ and ‘slow’ recycling, respectively \citep{RN100}. Either calcium chelation or stimulation with TNF$\alpha$ induced endocytosis of JAM-C in HUVECs \citep{RN515}. Subsequently, it was observed in tubular extensions from membranous structures formed at the cell junctions, but the endocytic pathway was not identified. The recruitment of JAM-C to the cell junctions depended on its interaction with junction-proximal scaffold proteins (likely ZO1 or -2), as indicated by an increase in its presence near the Golgi apparatus at the expense of the cell junctions upon deletion of its PDZ-binding motif. JAM-A endocytosis in Sertoli cells, which generate the hemato-testicular barrier, was blocked by knockdown of the clathrin heavy chain \citep{RN468}. 

The induction of macropinocytosis appears to be common to the responses of all tight junction integral proteins to inflammatory agonists: lipopolysaccharide (LSP) and CCL2 induced the also the micropinocytosis of JAM-A \citep{RN73}, but more recent studies revealed that under these conditions, JAM-A trafficking differed from claudin and occludin. The JAM-A population that translocated from the cell junctions and relocated to cytoplasmic punctae in immortalized mouse brain bEND.3 ECs was separate from internalized occludin and claudin-5, as well as from vascular endothelial (VE)-cadherin. JAM-A translocated from the cell junctions to the cytoplasm in 10-20 min and recycled back to the PM in 30-60 min, substantially faster than the 24-48-hour dynamics observed by previous studies \citep{RN100,RN101}. Internalized JAM-A collocated with and required the activity of Rab34, a known mediator of macropinosome formation \citep{RN341}. Subsequently, endocytosed JAM-A collocated with Rab5 and Rab4, indicating it trafficked to early endosomes and recycled rapidly \citep{RN73}. JAM-A was not collocated with Rab7, suggesting that only a small fraction of its population underwent lysosomal degradation. Similar to the response of JAM-A to IFN$\gamma$, lipopolysaccharide (LPS) and CCL2-induced macropinocytosis required the activities of RhoA and Rho-associated protein kinase (ROCK). 

The dynamics of JAM-C in quiescent human dermal microvascular ECs were dissimilar from JAM-A. In contrast to the latter, approximately 80 percent of the cellular JAM-C was in the cytoplasm \citep{RN342}. The remaining 20 percent were distributed diffusely on the cell surface rather than sequestered at the cell junctions. VEGF stimulation increased the cell-surface associated JAM-C fraction to 60 percent in approximately one hour. Interestingly, the translocation was anterograde, opposite to the VEGF-induced translocation of all other cell junction integral proteins, possibly paralleling the apparent promotion of intercellular permeability by JAM-C, contrary to JAM-A. A recent study on JAM-C dynamics in quiescent HUVECs reported that JAM-C was primarily at the junctions of confluent cells, and that two thirds of this population was removed from the junctions in two hours by constitutive recycling \citep{RN515}. The cytoplasmic JAM-C population was detected partially in early endosomes and in MVBs, indicating that it was targeted at least in part to lysosomal degradation. In essence, the effect of VEGF on JAM-C dynamics is opposite to its effect on JAM-A. Monolayer permeability increased despite the recruitment of JAM-C to the cell junctions. The mechanism that confers this intriguing effect is unknown. Mutation of all four lysines located in the JAM-C cytoplasmic domain to arginines, intended to abolish its ubiquitination, resulted in an increase in JAM-C abundance in early endosomes, coupled with its decrease in MVBs. Unlike wild type JAM-C, which was targeted by ubiquitination to MVBs and, subsequently, to lysosomes, the mutant did not associate with the E3 ligase Casitas B-lineage lymphoma (CBL), supporting CBL’s role in JAM-C recruitment to MVBs.
\\[2\baselineskip]
\begin{center}
\textbf{2.4. FUNCTIONAL CONSEQUENCES OF TIGHT JUNCTION PROTEIN TRAFFICKING}
\end{center}

In general, removal of integral proteins from cell junctions by endocytosis is detrimental to junction integrity. Their momentary abundance at the junction reflects a dynamic balance between endocytosis and the incorporation of newly synthesized or recycled proteins in the PM. Constitutive recycling in quiescent cells maintained the abundance of claudin-1, 4/5, and 16 at a steady state, thus preserving junction integrity \citep{RN1,RN233,RN236}. Depletion of Rab14, which regulates recycling, increased the abundance of claudin-2 at the cell junctions \citep{RN157} and intercellular junction integrity. Claudins form ion-selective pores \citep{RN533}, hence, the removal of claudin-2 from MDCK cell junctions reduced junction permeability to Na\textsuperscript{+}, whereas the depletion of claudin-4 and 7 reduced Cl\textsuperscript{-} permeability \citep{RN152}. Paracellular permeability to macromolecules was, however, invariably reduced upon increase in claudin abundance, though constitutive recycling rate varied among claudins \citep{RN1}. Occludin underwent constitutive recycling similar to that of the claudins, mediated by the same GTPase, Rab13 \citep{RN282,RN13}. In contrast to constitutive endocytosis, the endocytosis of claudins \citep{RN234,RN243,RN478,RN53}, occludin \citep{RN234,RN7,RN63,RN53,RN101}, and JAM-A \citep{RN100,RN515,RN342,RN73,RN101,RN468} induced by factors (EGF, VEGF, TGF$\beta$, TNF$\alpha$) or by cytokines (CCL2, IFN$\gamma$), increases cellular monolayer permeability. As recounted above, JAM-C is an intriguing exception to this rule \citep{RN515}.

\begin{center}
\textbf{3. MEMBRANE TRAFFICKING OF ADHERENS JUNCTION INTEGRAL PROTEINS}
\\[\baselineskip]
\textbf{3.1. CADHERINS}
\end{center}

Cadherin membrane trafficking is better understood than that of any other cell junction integral protein. Cadherins are the largest junction transmembrane protein family, consisting of 114 members \citep{RN78}. They adjoin neighboring cells by homophilic binding interactions. This review focuses on the most abundant (though not the only) cadherins in epithelial and in ECs: epithelial (E) and VE-cadherin, respectively. It includes relevant studies on neuronal (N) cadherin, which is expressed also in endothelial \citep{RN513} and epithelial \citep{RN514} cells. Human E- and VE-cadherin are 882 and 784 amino acid-long single-pass transmembrane proteins. They harbor five modular immunoglobulin-like extracellular-cadherin domains and cytoplasmic domains of 151 and 163 amino acids, respectively (Fig. 4). Unlike claudins, JAMs, and nectins, cadherins do not harbor a PDZ-binding motif and do not interact directly with PDZ domain-containing scaffold proteins. The canonical binding partners of their cytoplasmic domains are three members of the catenin protein family, $\alpha$-, $\beta$-, and $\delta$-catenin (p120) \citep{RN357}. $\beta$-catenin binds the carboxy-terminus region of the E-cadherin cytoplasmic domain (residues 811-882 of human E-cadherin) and $\alpha$-catenin \citep{RN359}, which crosslinks E-cadherin to f-actin under stretch \citep{RN361}. $\delta$-catenin binds the juxtamembrane region of the E-cadherin cytoplasmic domain (residues 758-769) \citep{RN360}. $\alpha$-catenin crosslinks E-cadherin to f-actin under stretch \citep{RN361}. VE-cadherin interacts with the three catenins in a similar manner \citep{RN362}. There is ample evidence that VE-cadherin endocytosis in response to VEGF reduces endothelial barrier integrity and increases its permeability \citep{RN4,RN6,RN520}.

\begin{figure*}[h]
\center
\includegraphics[scale=0.6]{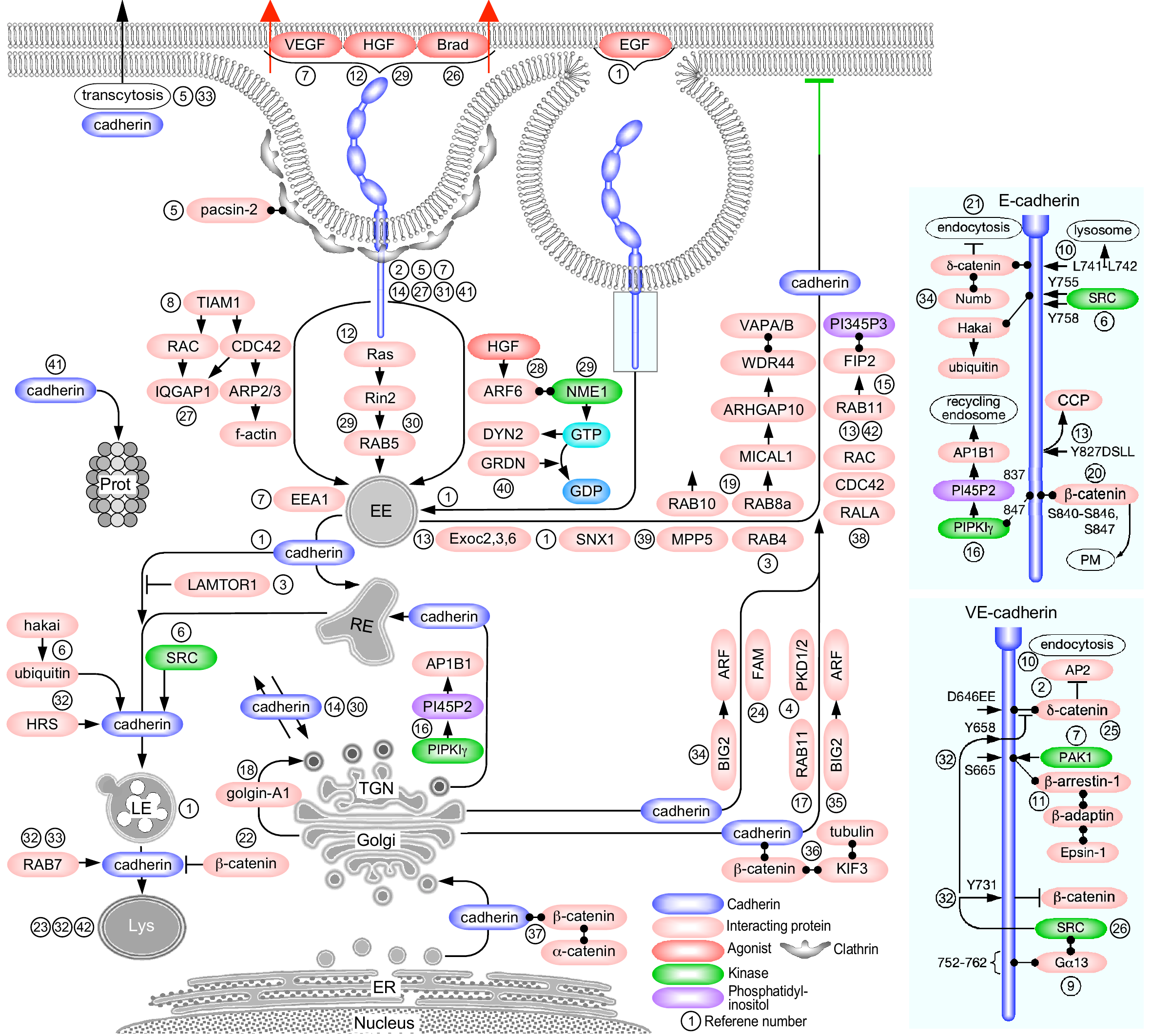}

\caption{\textbf{E- and VE-cadherin trafficking pathways} E- and VE-cadherin were internalized via clathrin-mediated endocytosis or via macropinocytosis. They were partially retrieved to the PM via Rab8a, Rab10, and Rab11-mediated recycling or diverted to lysosomal or proteasomal degradation. VEGF-induced endocytosis increased adherens junction permeability (red arrow); HGF-induced endocytosis increased adherens junction disassembly. Rab11a-mediated VE-cadherin recycling restored adherens junction integrity (green T). Inset shows phosphorylation sites and binding proteins to the cytoplasmic domain of each cadherin. Brad, bradykinin; CCP, clathrin-coated pit; EE, early endosome; GRDN, girdin; E, late endosome; Lys, lysosome; Prot, proteasome; RE, recycling endosome. Numbers correspond to the following references: (1) \citep{RN163}, (2) \citep{RN125}, (3) \citep{RN492}, (4) \citep{RN488}, (5) \citep{RN204}, (6) \citep{RN94}, (7) \citep{RN4}, (8) \citep{RN179}, (9) \citep{RN201}, (10) \citep{RN190}, (11) \citep{RN147}, (12) \citep{RN178}, (13) \citep{RN169}, (14) \citep{RN87}, (15) \citep{RN392}, (16) \citep{RN177}, (17) \citep{RN107}, (18) \citep{RN390}, (19) \citep{RN219}, (20) \citep{RN197}, (21) \citep{RN471}, (22) \citep{RN121}, (23) \citep{RN193}, (24) \citep{RN494}, (25) \citep{RN161}, (26) \citep{RN520}, (27) \citep{RN96}, (28) \citep{RN90}, (29) \citep{RN97}, (30) \citep{RN116}, (31) \citep{RN98}, (32) \citep{RN473}, (33) \citep{RN135}, (34) \citep{RN129}, (35) \citep{RN485}, (36) \citep{RN108}, (37) \citep{RN114}, (38) \citep{RN186}, (39) \citep{RN137}, (40) \citep{RN137}, (41) \citep{RN20}, (42) \citep{RN25}.}

\vspace{-0.5cm}

\end{figure*}

\textbf{3.1.1. Endocytosis}

Similar to tight junction transmembrane proteins, E-cadherin was construed to undergo constitutive endocytosis, induced by transferring either MDCK cells \citep{RN87}, or MCF-7 epithelial breast cancer cells \citep{RN98} from 18$\circ$C to 37$\circ$C. This conclusion was based on the inhibitory effectiveness of K\textsuperscript{+} depletion \citep{RN377} and of the pharmacological inhibitor bafilomycin A1 \citep{RN279}. The presence of a YDSLL motif at position 827 of human E-cadherin cytoplasmic domain, known to recruit the host protein to clathrin-coated pits \citep{RN86}, further supported this premise \citep{RN87}. In contrast, the E-cadherin constitutive endocytosis in MCF-7 cells was found to be clathrin-independent. Nevertheless, E-cadherin was detected in clathrin-coated pits by transmitted electron microscopy (EM), and, as the authors admitted, clathrin-dependent endocytosis may have occurred before the earliest sampled time point of 5 min post the 118$\circ$C to 37$\circ$C switch \citep{RN98}. Constitutive endocytosis of VE-cadherin in human dermal microvascular ECs was not positively identified, but its attributes were consistent with clathrin-dependent endocytosis \citep{RN110}.

VE-cadherin in VEGF-treated HUVECs appeared in cytoplasmic vesicles as soon as 2 min after VEGF administration, where it collocated with clathrin, dynamin-2, and the early endosome markers Rab5 and EEA1 \citep{RN4}. The cytoplasmic domain VE-cadherin harbors a conserved region close to the $\delta$-catenin binding site that was phosphorylated by p21-activated kinase (PAK)-1 on S665 in a Src-dependent manner. The phosphorylation of this site was required and sufficient for VE-cadherin endocytosis. In contrast, a non-phosphorylatable S665V mutant remained at the cell junctions, whereas a S665D phosphomimetic mutant was endocytosed constitutively in the absence of VEGF \citep{RN4}. The phosphorylation of S665 likely recruited the clathrin-binding endocytic adapter arrestin-$\beta$1, which bound to and collocated with endocytosed VE-cadherin. Src is constitutively active in vein ECs, inducing the phosphorylation of Y658 and Y685 in the cytoplasmic domain of VE-cadherin in HUVECs \citep{RN520}. The phosphorylation sensitized VE-cadherin to bradykinin, an inflammatory cytokine that increases vessel permeability \citep{RN519}. Bradykinin induced VE-cadherin ubiquitination and clathrin-mediated endocytosis, without dissociating $\delta$-catenin. Apparently, the promotion of endocytosis by ubiquitination overcame the inhibitory effect of $\delta$-catenin binding to VE-cadherin.

VE-cadherin was tyrosine-phosphorylated in resting mouse tracheal vein (but not in artery) ECs \citep{RN520}. The phosphorylation, which was localized to Y658 and Y685 in the VE-cadherin cytoplasmic domain, was attributed to Src, having been abolished by a specific pharmacological inhibitor, AZD0530 \citep{RN532}. Bradykinin, an inflammatory cytokine that increases vessel permeability \citep{RN519}, induced clathrin-mediated endocytosis and ubiquitination of tyrosine-phosphorylated VE-cadherin. The endocytosed VE-cadherin remained bound to $\delta$-catenin, suggesting that the promotion of endocytosis by ubiquitination overcame the inhibitory effect of $\delta$-catenin binding to VE-cadherin. Both effects were blocked by Src inhibition or by replacement of Y658 and Y685 by phenylalanine, indicating that endocytosis and ubiquitination required phosphorylation of VE-cadherin. Exposure of HUVECs to flow produced the same tyrosine phosphorylation, suggesting that the phosphorylation of VE-cadherin in vein ECs was a hemodynamic effect.

A recent study reported that the proteoglycan syndecan-4 collocated with VE-cadherin along HUVEC junctions and interacted with it independently of the former’s glycosaminoglycan chains \citep{RN523}. VEGF administration induced Src-mediated phosphorylation of Y180 in the syndecan-4 cytoplasmic domain, followed by co-endocytosis with VE-cadherin. VEGF-induced endocytosis of VE-cadherin in ECs devoid of syndecan-4 was approximately 3-fold lower and the increase in their permeability was around half than in syndecan-4-expressing ECs. The manner by which syndecan-4 facilitated VE-cadherin endocytosis was not identified.

An insight into the relation between the homophilic \textit{trans} interaction of cadherin ectoplasmic domains and cadherin endocytosis into its host cell was provided by coculturing A-431 human epidermal carcinoma epithelial cells that expressed E-cadherin with either a W213C or T227C mutation \citep{RN111}. The positions of the cysteines permitted only the crosslinking of W213C-T227C \textit{trans} E-cadherin dimers once dimerization was induced by elevating Ca\textsuperscript{2+} concentration to 1 mM and reducing the respective sulfhydryl groups of each mutant \citep{RN370} with the cysteine-specific crosslinker DPDPB \citep{RN507}. The relation between E-cadherin binding in \textit{trans} and its endocytosis was detected by tracking the abundance of crosslinked heterodimers by immunoprecipitation. ATP depletion by sodium azide or hypertonic shock produced by medium supplementation with sucrose, conditions known to inhibit endocytosis \citep{RN376,RN371} prevented intercellular junction disruption by calcium chelation or by trypsin treatment. These results suggested that endocytosis was the main mechanism of adhesive E-cadherin dissociation \citep{RN111}. In contrast, f-actin disruption by cytochalasin D or by latrunculin A, or reduction of myosin-generated tension by the ROCK inhibitor Y-27632 or the MLCK inhibitor ML-7, had minor effects on the abundance of E-cadherin dimers or on their formation rate. EM analysis showed that E-cadherin was internalized in endocytic vesicles emanating from adherens junctions (ibid.).

Observations made by two-photon fluorescence recovery after photobleaching (FRAP) and fast 3D fluorescence microscopy determined less than 10 percent of the total cell-surface E-cadherin in MDCK cells was free to diffuse laterally \citep{RN167}. The majority of the population recycled between binding in \textit{trans} at the cell surface and the cytoplasm by endocytosis, with a surface residence time of 4 min. Close to half of the cytoplasmic E-cadherin translocated from the PM by endocytosis within less than 3 min. In support of endocytosis as an exchange mechanism between the trans-interacting and cytoplasmic E-cadherin populations, both dynasore and myristyl trimethyl ammonium bromide (MiTMAB), a noncompetitive inhibitor of the GTPase activity of dynamin \citep{RN380}, suppressed E-cadherin exchange between the cytoplasm and the PM. The endocytosis rate and its inhibitor sensitivity suggested that it was clathrin-mediated. Subsequent studies measured the turnover of E-cadherin in quiescent A-431 cells by fusing its cytoplasmic domain to the green-to-red photoconvertible protein Dendra2 \citep{RN190}. This technique distinguished between newly recruited E-cadherins fluorescing in green and red-fluorescing E-cadherins that dissociated from cell junctions after photoconversion. Junction-residing E-cadherins were replaced continuously with a half-time of approximately 2.5 min. E-cadherin turnover at the cell junctions was apparently energy-dependent, because ATP depletion slowed it substantially.  

The activity of the ADP-ribosylation factor (Arf)-6, a GTPase that regulates membrane trafficking between the PM and recycling endosomes \citep{RN373}, was essential for E-cadherin endocytosis. Expression of an Arf6 dominant-negative mutant blocked hepatocyte growth factor (HGF)-induced internalization of E-cadherin downstream of Src \citep{RN90}. A subsequent study determined that Arf6 was required for dynamin activation \citep{RN97}. Yeast-two-hybrid screen of Arf6-binding proteins identified nucleoside diphosphate kinase A (NME1), which synthesizes the GTP that drives dynamin activity \citep{RN366}. E-cadherin endocytosis in MDCK cells entailed the recruitment of NME1 to GTP-Arf6 \citep{RN97} and sequestration of the Rac1 guanine exchange factor (GEF) TIAM1 \citep{RN96}. The ensuing reduction of Rac1 activity at the cell junctions presumably facilitated the inward translocation of endocytosed vesicles. The expression of dominant-negative dynamin 2-K44A hindered E-cadherin endocytosis. Because E-cadherin did not collocate with caveolin, its endocytosis was apparently clathrin-dependent. Induction of vesicle budding in an adherens junction-enriched fraction of rat liver cells followed by $\beta$-catenin vesicle immunoisolation, revealed that E-cadherin co-sedimented with the clathrin heavy chain, epsin-1, and the clathrin-coated vesicle component $\alpha$-adaptin (AP2B1) (Brodsky \citep{RN367,RN147}. Further establishing the dependence of E-cadherin endocytosis on clathrin, expression of the epsin N-terminal homology (ENTH) domain, which inhibits clathrin-dependent endocytosis of the EGF and insulin receptors \citep{RN145}, blocked either HGF or low calcium-induced E-cadherin endocytosis in MDCK cells \citep{RN147}. A chimera consisting of the E-cadherin ectoplasmic domain and the Fc region of human IgG expressed in mouse fibroblasts was endocytosed at 2 $\mu$M but not at 2 mM calcium, indicating that only non-\textit{trans}-interacting E-cadherin on the surface of adjoining cells was endocytosed, in agreement with Troyanovksy et al. \citep{RN111}. \textit{Trans}-bound E-cadherin induced the activation of the Rho GTPases Rac and Cdc42 \citep{RN147}. In turn, active Rac and Cdc42 inhibited E-cadherin endocytosis, presumably by the f-actin cross linking activity of their effector, ras GTPase-activating-like protein IQGAP1 \citep{RN369,RN368}. In agreement, f-actin depolymerization by latrunculin-A augmented E-cadherin endocytosis. The scission of E-cadherin-endocytic vesicles in HeLa cells by dynamin-2 required the GAP activity of girdin, an actin and $\alpha$-adaptin-binding protein, which interacts specifically with GTP-bound dynamin-2 and catalyzes its hydrolysis \citep{RN137}. The E-cadherin sorting mechanism to girdin-associated coated pits was not reported. The interaction of E-cadherin with $\alpha$-adaptin, and, indirectly, with clathrin \citep{RN109} was regulated in MCF-7 epithelial cells by Numb, via its binding to the carboxy-terminus of $\delta$-catenin \citep{RN129}. The E-cadherin interaction with Numb was required for its endocytosis and targeting to the cell’s basolateral region. The phosphorylation of Numb by PKC$\zeta$ inhibited its binding to E-cadherin and $\alpha$-adaptin, but the manner by which PKC$\zeta$ activity was regulated was not specified.

E-cadherin endocytosis was linked to another adherens junction transmembrane protein, nectin. The endocytosis of non-cross-interacting E-cadherin expressed in the aforementioned mouse fibroblasts \citep{RN147} was reduced significantly when these cells were transfected by full-length nectin-1, but not by nectin-1 devoid of the four amino-acid PDZ-binding domain at the carboxy terminus of its cytoplasmic domain \citep{RN152}, presumably because the latter was unable to bind afadin. While the expression of full-length afadin in fibroblasts expressing E-cadherin and nectin-1 did not inhibit E-cadherin endocytosis, the expression of afadin lacking the amino-terminus domain reduced endocytosis, apparently because that domain sequesters Rap1\citep{RN152}. The role of Rap1 in E-cadherin endocytosis was deduced to be the stabilization of E-cadherin binding to $\delta$-catenin, because co-expression of constitutively active Rap1 with E-cadherin and nectin-1 increased the abundance of E-cadherin-coimmunoprecipitated $\delta$-catenin compared to cells that expressed nectin-1 devoid of the PDZ-binding motif. The Rap1-mediated stabilization mechanism was not specified.

The incorporation of epithelial cell surface proteins that harbor a dileucine motif, including E-cadherin, into clathrin-coated vesicles requires interaction with the AP1 complex subunit $\beta$1 (AP1B1) \citep{RN375}. Type I$\gamma$ phosphatidylinositol phosphate kinase (PIPKI$\gamma$) synthesizes phosphatidylinositol (4,5) phosphate (PtdIns(4,5)P\textsubscript2), a phosphoinositol that facilitates the assembly of coat protein complexes \citep{RN374}. PIPKI$\gamma$ associated preferentially with dimeric E-cadherin in MDCK cells, as well as with VE-cadherin in HUVECs and with N-cadherin in HEK-293 cells \citep{RN177}. The E-cadherin binding region was narrowed down to amino-acids 837-847 in its cytoplasmic domain, a motif conserved in VE- and N-cadherin. Interference with PIPKI$\gamma$ activity by either knockdown or expression of a kinase-dead mutant resulted in depletion of E-cadherin from the cell junctions and its accumulation in cytoplasmic vesicles. In addition to its enzymatic activity, PIPKI$\gamma$ recruited E-cadherin to endocytic vesicles by binding to subunit $\beta$1 of the AP1 complex. Replacement of the dileucine motif of E-cadherin in MCDK cells (L744-L745, corresponding to L741-L742 of human E-cadherin) with alanines reduced substantially E-cadherin endocytosis in cells transferred from 4$\circ$C to 37$\circ$C \citep{RN121}. Whereas knockdown of $\delta$-catenin increased the endocytosis of wild type E-cadherin, it had no such effect on the leucine to alanine mutant. Apparently, both $\delta$-catenin dissociation and the presence of the dileucine motif are required for E-cadherin endocytosis.

The role of both the conserved dileucine motif and a presumably ubiquitinated K738 by in the endocytic sorting of cadherin was confirmed by their mutation to valine-alanine and to arginine, respectively \citep{RN190}. Consequently, the constitutive endocytosis of the E-cadherin mutant in quiescent A-431 cells was practically blocked. Surprisingly, the mutation had no effect on E-cadherin turnover at the cell junctions, indicating that endocytosis cannot fully account for it. The turnover mechanism was not identified. Further analysis revealed that the binding of $\delta$-catenin to the cytoplasmic domain of E-cadherin and the interaction of the dileucine motif with clathrin adaptor complexes were not entirely independent of each other \citep{RN126}. $\delta$-catenin binding involved two regions in the cytoplasmic domain, a ‘static’ motif spanning residues 747-781, and a ‘dynamic’ site that encompasses the dileucine endocytosis motif. The latter site was permissive to binding competition between $\delta$-catenin and clathrin adaptor complexes such as AP2, whose binding would initiate E-cadherin endocytosis. $\delta$-catenin-d binding to E-cadherin was regulated by phosphorylation: MAPK1/3 (ERK)-mediated phosphorylation of T310 of $\delta$-catenin in MCF10A cells undergoing wound closure abolished its binding to E-cadherin and released the latter from the cell surface \citep{RN164}. Glycogen synthase kinase-3 (GSK3)-mediated phosphorylation of the same residue dissociated $\delta$-catenin from N-cadherin in astrocyte cells \citep{RN191}. 

The association of E- or VE-cadherin with the endocytic machinery was mediated by their cytoplasmic domains. An internalized chimera consisting of the ectoplasmic and transmembrane domains of the interleukin (IL) 2 receptor (IL2R) and the cytoplasmic domain of VE-cadherin collocated extensively with endogenous VE-cadherin in early endosomes when constitutive endocytosis was induced by transferring human dermal microvascular cells from 4$\circ$C to 37$\circ$C \citep{RN110}. Cytosol acidification and K\textsuperscript{+} depletion, approaches shown to inhibit clathrin endocytosis \citep{RN376,RN377}, inhibited the endocytosis of the IL2R-VE-cadherin chimera. The chimera internalized in 5 min after transference to 37ºC and collocated with endocytosed transferrin, a marker of clathrin-dependent endocytosis \citep{RN378}. Overexpression of $\delta$-catenin blocked internalization of IL2R-VE-cadherin, whereas replacing the EMD binding motif at position 652 of human VE-cadherin \citep{RN160} to alanines abolished $\delta$-catenin binding. Chimeric VE-cadherin collocated with clathrin and appeared in cytoplasmic punctae \citep{RN20}. These results indicated that the clathrin-mediated endocytosis of VE-cadherin requires dissociation of $\delta$-catenin from its cytoplasmic domain. Once endocytosed, both $\delta$- and $\beta$-catenin dissociated from VE-cadherin. Subsequent studies determined that $\delta$-catenin blocked VE-cadherin endocytosis by sequestering it away from membrane domains that were presumably populated by the clathrin-adaptor complex AP2 via an unknown mechanism \citep{RN125}. In contrast to Xiao et al. \citep{RN20}, replacement of the EMD motif to alanines resulted only in a small reduction of the mutant’s endocytosis in A-431 cells \citep{RN161}. Rather, replacement of the GGG motif at position 649 in the $\delta$-catenin-binding region of human VE-cadherin abolished the binding of $\delta$-catenin almost completely. The $\delta$-catenin binding region in human and mouse VE-cadherin spanned residues K627 – N664, out of which the K627 – S634 sub-region corresponds to the aforementioned E-cadherin ‘dynamic’ binding site \citep{RN126}, and the remaining V635 – N664 sub-region corresponds to the ‘static’ binding site. $\delta$-catenin binding to VE-cadherin inhibited its endocytosis by masking the DEE motif at position 646 and prevented the binding of Kaposi sarcoma virus ubiquitin ligase K5 to K626 and K633. A subsequent study from the same group revealed that the 646DEE motif was sufficient for VE-cadherin endocytosis, whereas the abolishment of $\delta$-catenin binding alone by mutating 649GGG to AAA was not \citep{RN521}. The role of the DEE motif in VE-cadherin endocytosis was not specified. 

E-cadherin endocytosis required the activity of Rho GTPases and their effects on the actin cytoskeleton \citep{RN383}. Expression of a Cdc42 loss-of-function (LOF) mutant in the epithelial cells of the Drosophila dorsal thorax resulted in a drastic reduction in the number of E-cadherin vesicular structures in the cytoplasm \citep{RN179}, and in the appearance of tubular endocytic vesicles associated with adherens junctions \citep{RN168}. A LOF mutation in actin-related protein (Arp)-3, a subunit of the Arp2/3 complex and a Cdc42 effector \citep{RN384}, resulted in similar blockage of E-cadherin endocytosis \citep{RN179}. LOF mutations of Par6 or of p21-activated kinase (Pak)-1, both of which are Cdc42 effectors \citep{RN388,RN387}, phenocopied the effect of Cdc42 LOF mutation on E-cadherin endocytosis \citep{RN168}. In vitro experiments in HUVECs and in vivo experiments in zebrafish revealed that homophilic VE-cadherin \textit{trans} dimers on adjacent cells underwent transcytosis while remaining bound to $\delta$-catenin \citep{RN135}, in agreement with previous studies of endocytosis in vivo \citep{RN520}, but contrary to in vitro studies \citep{RN20}. The contradicting results may reflect differences between endocytosis in \textit{cis} versus in \textit{trans}, or between in vitro versus in vivo conditions. The transcytosis required myosin-driven traction transmitted via f-actin, and Rac1 activity. The endocytic pathway was not specified, but none of the known pathways appeared to be involved. A contrasting phenomenon was observed in motile HUVECs when they formed transient focal adherenes junctions that anchored force-bearing stress-fibers \citep{RN203}. VE-cadherin located in these junctions was subjected to imbalanced tenstile forces that generates asymmetric junction geometry. Despite the asymmetry, only 9 percent of the VE-cadherin cellular population underwent transcytosis, whereas the majority was endocytosed into its host cell \citep{RN204}. The endocytosis was apparently balanced out by recruitment of the bin, amphiphysin, and rvs (BAR) domain protein pacsin-2 to the curved PM around focal adherens junctions. Endocytosed VE-cadherin was detected in Rab5 and Rab4-labeled vesicles, indicating that it recycled rapidly to the PM. Similar to tight junction transmembrane proteins, E-cadherin was internalized by macropinocytosis. The application of EGF to MCF-7 epithelial breast cancer cells resulted in the collocation of E-cadherin with $\delta$ and $\beta$-catenin in EEA1-associated 2-2.25 $\mu$m macropinosomes, followed by translocation to the juxtanuclear region \citep{RN163}. Part of this population was in late endosomes and lysosomes, but its majority was recycled back to the cell surface (see below).

E-cadherin endocytosis was associated with several types of posttranlational modifications. Viral Src expressed in MDCK cells phopshorylaetd Tyr755 and 758 in the cytoplasmic domain of E-cadherin \citep{RN94}. The phosphorylation enabled binding of the E3 ligase hakai and ubiquitination of E-cadherin, increasing the endocytosis of the latter in response to HGF. More recent studies \citep{RN201} addressed the Src-mediated phosphorylation of Tyr658 of VE-cadherin in human lung microvascular ECs, a modification known to dissociate $\delta$-catenin from VE-cadherin and facilitate its endocytosis \citep{RN473}, was triggered by trimeric G$\alpha$13 GTPase. G$\alpha$13 binds both Src and residues 752-762 of the VE-cadherin cytoplasmic domain \citep{RN201}. E-cadherin expressed in A-431 cells was constitutively phosphorylated on serines 840, 846 and 847 in the $\beta$-catenin binding motif \citep{RN201}. The phosphorylation facilitated $\beta$-catenin binding and favored the recruitment of E-cadherin to the PM. The kinase was not identified but was presumed to reside in the Golgi apparatus. Contrary to the latter study, Src-mediated phosphorylation of VE-cadherin on Tyr731 in HUVECs prevented $\beta$-catenin binding, and, consequently, would have favored VE-cadherin endocytosis and lysosomal targeting \citep{RN473}.  

Endocytosed E-cadherin in MDCK and MCF-7 cells and endocytosed VE-cadherin in human dermal microvascular ECs underwent Rab5-mediated trafficking to early endosomes \citep{RN178,RN116}. In response to HGF, Rab5 was activated by the GEF Ras and Rab interactor (RIN)-2 downstream of Ras \citep{RN178}. N-cadherin expressed in immortalized rat fibroblasts collocated with $\delta$-catenin in cytoplasmic vesicles that translocated retrogradely at 0.5-1.0 $\mu$m/s upon adherens junction disassembly induced by calcium chelation \citep{RN118}. Vesicle movement was disrupted by nocodazole, indicating they translocated along microtubules. The movement was driven by kinesin, recruited to the vesicular N-cadherin via $\delta$-catenin, which binds the kinesin heavy chain \citep{RN162}. 

\textbf{3.1.2. Recycling}

Multiple studies identified the Golgi apparatus and the TGN as major destinations or departure points of cadherin during either constitutive recycling \citep{RN87,RN98}, or as a result of the expression of dominant-negative Rac1 \citep{RN119}. Newly synthesized N-cadherin precursor expressed in HeLa cells was present in the ER and the Golgi apparatus prior to the cleavage of its signal peptide (residues 1-25) and propeptide (residues 26-159) \citep{RN114}. The carboxy-terminus of the N-cadherin precursor was phosphorylated and bound to $\delta$-catenin, followed by the biding of $\alpha$ and $\beta$ catenins as a dimer. The stoichiometric ratios of N-cadherin to the three catenins were close to unity, suggesting that most of the precursor N-cadherin was bound to catenins before it translocated from the ER and Golgi. Contrary to Wahl et al. \citep{RN114}, Miranda et al. ruled out systematically the presence of $\delta$-catenin in the ER and the Golgi apparatus of MDCK cells \citep{RN471}. The disagreement was attributed to unspecified phenotypic differences between HeLa and MDCK cells. The transport of N-cadherin from the Golgi apparatus to the periphery of mouse neural progenitor cells was driven by kinesin KIF3  \citep{RN108}, a microtubule plus end-directed anterograde molecular motor that associates with N-cadherin via $\beta$-catenin \citep{RN109}. Mutation of the dileucine motif at residues 741-742 of human E-cadherin expressed in MDCK cells interfered with its routing to the basolateral membrane, showing that the association with $\delta$-catenin was required for its correct targeting \citep{RN471}. This conclusion was challenged by a subsequent study, which found that E-cadherin mutants that lacked the whole cytoplasmic domain, or whose leucines were substituted by alanines, were detected exclusively on the PM of MDCK cells \citep{RN121}. Instead of the basolateral membrane, the latter study concluded that the dileucine motif targeted E-cadherin to lysosomes \citep{RN193} (see below). No explanation was provided to reconcile the disagreement between the conclusions of Miranda et al. and Miyashita and Ozawa in regard to the role of the dileucine motif in E-cadherin basolateral targeting.

Newly synthesized wild type E-cadherin, or a truncated mutant devoid of the dileucine motif expressed in the HeLa cells, translocated constitutively from the TGN via tubular or spherical endosomes that moved at an approximate rate of 3 $\mu$m/s \citep{RN390}. In a minority of cases, the vesicles reached the cell periphery in 40 s. After remaining stationary for 10 s, the vesicles disappeared, possibly after fusion with the PM. The initial membrane trafficking of E-cadherin away from the Golgi apparatus was shared with occludin and claudin-5 \citep{RN112}. The exit from the TGN was mediated by golgin-A1, a protein that binds directly to the TGN membrane \citep{RN390}. Golgin-A1 was required up to but not after the budding of E-cadherin-associated vesicle from the TGN. The recruitment of E-cadherin to the PM in MDCK cells required the activities of Rac1 and Cdc42 \citep{RN119}. Expression of dominant-negative mutants of the two GTPases resulted in the accumulation of E-cadherin in large perinuclear vesicles that collocated poorly with Golgi and TGN markers. The GTPase RalA that is a component of the exocyst and is required for its structural stability \citep{RN391}, enhanced the delivery of E-cadherin-containing vesicles from the TGN to the PM \citep{RN186}. Ubiquinated E-cadherin that would be normally destined to lysosomal degradation \citep{RN493} was diverted in subcobfluent T84 epithelial cells to recycling together with $\beta$-catenin by the deubiquitinase fat facets in mammals (FAM) \citep{RN494}. FAM collocated with both proteins in the Golgi apparatus and in cytoplasmic vesicles that were possibly headed to the PM. Because FAM interacted with E-cadherin and $\beta$-catenin only in subconfluent cells, this mechanism is apparently inactive once cell junctions are stabilized. 

Observations on epithelial proximal tubule cells from individuals carrying autosomal dominant polycystic kidney disease caused by LOF mutations in PKD1 and PKD2 revealed that E-cadherin was sequestered in perinuclear vesicles and was absent from the cell junctions \citep{RN488}. The mutant genes encode polycystins 1 and 2, large 11-pass trasmembrane proteins that are present in adherens junctions \citep{RN489} and bind E-cadherin \citep{RN490}. The nature of the vesicles and the manner by which polycistins facilitate the recruitment of E-cadherin to cell junctions are unknown. Another mechanism that favored E-cadherin recycling over degrdation was attributed to late endosomal/lysosomal adaptor and MAPK and MTOR activator (LAMTOR)-1 (named alternatively lipid raft adaptor protein p18), a protein that binds endosomal outer membranes and suppresses trafficking to lysosomes \citep{RN491}. When rat lung microvascular ECs were challenged by LPS, overexpression of LAMTOR1 suppressed the removal of VE-cadherin from the cell junctions and decreased its abundance in late endosomes in favor of early endosomes \citep{RN492}. These effects required LAMTOR1 binding to endosomes, and may have been exerted by the suppression of Rab7 and enhancement of Rab11 activities.

The involvement of Rab GTPases in cadherin membrane trafficking is known in more detail than for any other junction transmembrane protein. E-cadherin that underwent constitutive endocytosis in MCF-7 cells collocated with Rab11 and relocated back to the cell surface in 15-30 min \citep{RN98}. The majority of the above-mentioned vesicular E-cadherin expressed in HeLa cells \citep{RN390} trafficked in Rab11-associated recycling endosomes from which they moved to and fused with the PM. The expression of a dominant negative Rab11 mutant in MDCK cells resulted in mistargeting of E-cadherin which translocated to the apical instead of to the basolateral region \citep{RN196}. The recycling of VE-cadherin in human lung microvascular ECs required specifically the activity of Rab11a \citep{RN25}. VE-cadherin endocytosed in response to calcium chelation collocated with Rab11a in cytoplasmic vesicles. Depletion of Rab11a blocked VE-cadherin recycling back to the PM, resulting in its accumulation in lysosomes. VE-cadherin and Rab11a associated with each other by forming a ternary complex with Rab11 family-interacting protein (FIP)-2, a Rab11 effector that binds PtdIns(3,4,5)P\textsubscript3 and targets recycling vesicles to the PM \citep{RN392}. Knockdown of Rab11a impaired VE-cadherin recycling to the cell junctions and increased the vascular leakage in LPS-treated mice \citep{RN25}. Similar to its endocytosis from the PM, trafficking of E-cadherin from the TGN to recycling endosome required PIPKI$\gamma$ activity \citep{RN177}. The inhibition of brefeldin A-inhibited GEF (BIG)-2, an Arf GEF citep{RN487}, blocked the trafficking of E-cadherin and $\beta$-catenin from the TGN to the adherens junctions of MDCK cells \citep{RN485}. The role of Arf GTPases in this leg of E-cadherin trafficking could be the targeting of recycling vesicles to the PM \citep{RN373}. 

The trafficking of E-cadherin from recycling endosomes to the PM of MDCK cells depended on Rab8 and MICALL2 \citep{RN102}, a Rab8 and Rab13 effector involved in the recycling of tight junction transmembrane proteins \citep{RN13}. MICALL2 mediates membrane trafficking by regulating the assembly state of f-actin \citep{RN284}. A recent study found that GTP-bound Rab8a and Rab10 were recruited to elongated E-cadherin-containing tubular vesicles in the perinuclear region of HeLa cells \citep{RN219}, followed by the recruitment of MICAL1, an actin filament-severing protein \citep{RN220} and a likely Rab8 effector \citep{RN472}. In turn, MICAL1 recruited the Rho GTPase-activating protein 10 (ARHGAP10), a BAR domain containing protein, which extends and stabilizes tubular vesicles. ARHGAP10 recruited tryptophan-aspartate (WD) repeat-containing protein 44 (WDR44), a scaffold protein that bridges elongating tubules with the endoplasmic reticulum (ER) by binding to vesicle-associated membrane protein-associated proteins A and B (VAPA/B) \citep{RN221}. This mode of exocytosis suggests that E-cadherin segregated from other cargo proteins prior to exiting the ER \citep{RN219}.

The trafficking of E-cadherin in Drosophila epithelial cells from recycling endosomes to adherens junctions was mediated by the exocyst octameric complex. LOF of the genes encoding the Exoc2, Exoc3, and Exoc6 subunits, which are required for vesicle tethering to the PM \citep{RN393}, resulted in the accumulation of VE-cadherin, and of $\alpha$ and of $\beta$-catenin in enlarged Rab11-marked recycling endosomes \citep{RN169}. E-cadherin shares with claudins the syntaxin-4-dependent mechanism of docking to the PM described above \citep{RN14}. E-cadherin delivery to the PM in MDCK cells required an apicobasal complex protein, protein associated associated with lin seven (PALS1)-1 (named alternatively MPP5). In its absence, E-cadherin accumulated in cytoplasmic punctae adjacent to the cell periphery that did not collocate with early or recycling endosomes, or with Golgi markers \citep{RN173}. The accumulation was caused probably by mislocation of the exocyst, as was indicated by the absence of Exoc4 from the cell junctions. The recycling of E-cadherin that underwent macropinocytosis into MCF-7 cells was mediated by sorting nexin (Snx)-1 \citep{RN163}, an adaptor protein that mediates cargo selection by the retromer complex and retrograde trafficking to the TGN \citep{RN389}. Given its involvement in early endosome to Golgi retrieval of proteins destined for degradation \citep{RN497}, association with Snx1 was expected to divert E-cadherin from degradation to recycling \citep{RN163}. In agreement, Snx1 knockdown reduced substantially the recycling of E-cadherin to the cell surface.

\textbf{3.1.3. Degradation}

Inhibition of the proteasome by MG132 blocked VE-cadherin constitutive endocytosis in dermal microvascular ECs altogether, whereas inhibition of lysosome activity by chloroquine resulted in the accumulation of VE-cadherin in cytoplasmic punctae \citep{RN110}. These results suggested that both degradative pathways were involved in the processing of VE-cadherin, albeit at different stages of endocytosis. The manner by which catalytic inhibition of proteasome activity by MG132 \citep{RN516} blocks VE-cadherin endocytosis is unknown.

Temperature-dependent overexpression of constitutively active Src in MDCK cells increased substantially the preexisting basal abundance of E-cadherin lysosomal digestion products at the permissive temperature \citep{RN116}. E-cadherin was shown to be a Src substrate, suggesting that its tyrosine phosphorylation induced ubiquitination, possibly by the hakai ligase \citep{RN94}, followed by targeting to lysosomes. The lysosomal sorting was mediated by the endocytic scaffold protein HRS, which collocated with E-cadherin in cytoplasmic vesicles and in LAMP1-associated endosomes. The switch to permissive temperature enhanced Rab7 activity, whereas the expression of dominant-negative Rab7 blocked E-cadherin lysosomal degradation, in agreement with the established role of Rab7 in trafficking to lysosomes \citep{RN496}. The dependence of E-cadherin targeting to the lysosome on Rab7 was shared by VE-cadherin undergoing constitutive endocytosis in HUVECs \citep{RN135}.

Aside from its role in regulating cadherin endocytosis, the dileucine motif was implicated in the targeting of cadherin to the lysosome (Miyashita and Ozawa, 2007a). This conclusion stemmed from observations that replacement of the leucines with alanines resulted in the recruitment of E-cadherin to the basolateral membrane of MDCK cells and its absence from the lysosome, unlike wild-type E-cadherin. Dileucine motif sorting of E-cadherin required $\beta$-catenin binding to the cadherin cytoplasmic domain \citep{RN193}, though the binding region of the latter is close to 70 amino acids downstream of the motif. E-cadherin mutants that lost the ability to bind $\beta$-catenin translocated through the biosynthetic pathway and reached early endosomes, but instead of undergoing exocytosis to the PM, they were diverted to lysosomes \citep{RN193}. It was hypothesized that $\beta$-catenin binding to the natively folded E-cadherin cytoplasmic domain blocked the dileucine motif, thus preventing E-cadherin recruitment to lysosomes. 

\begin{center}
\textbf{3.2. NECTINS}
\end{center}

\begin{figure*}[h]
\center
\includegraphics[scale=0.60]{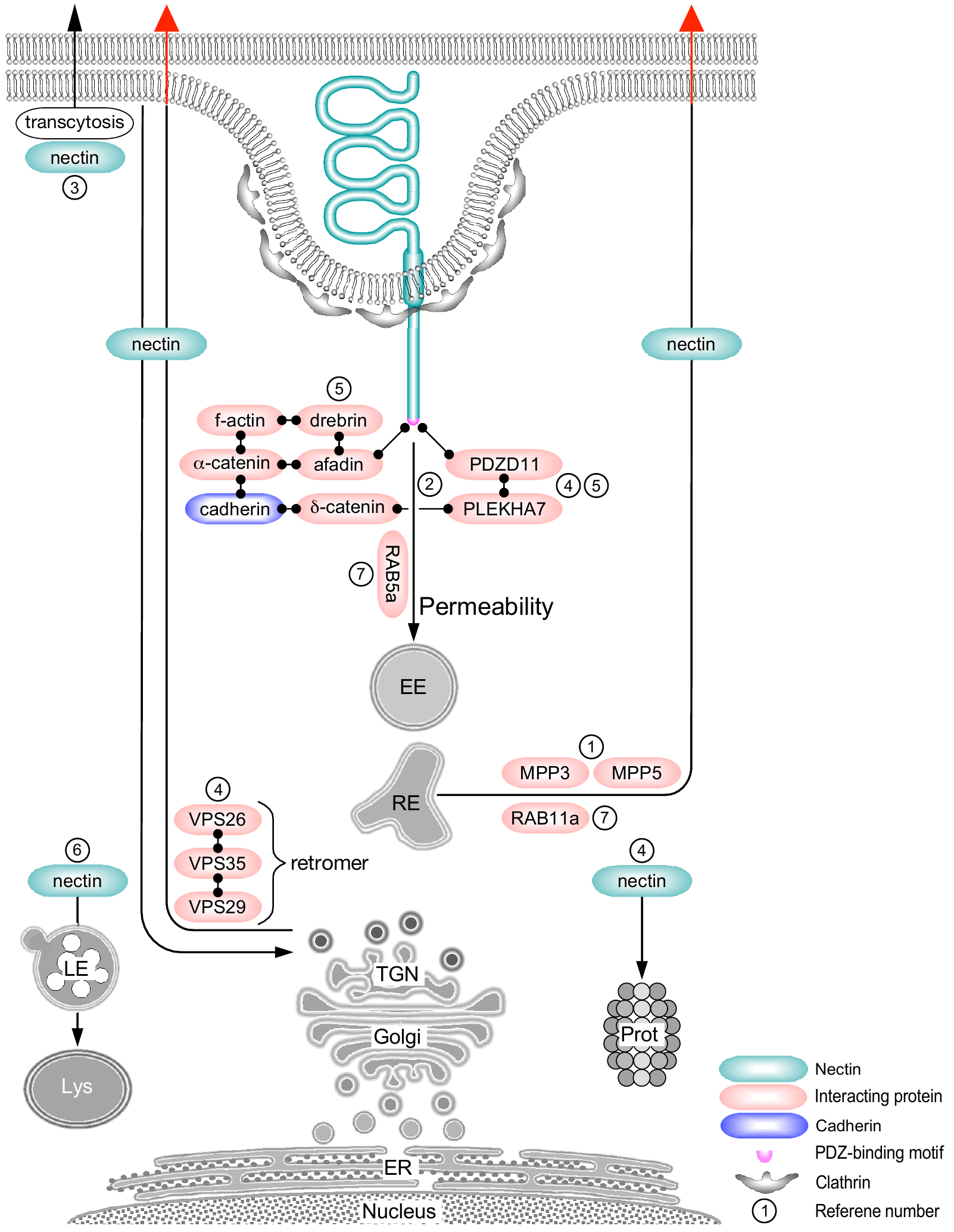}
\vspace{-0.0cm}

\caption{\textbf{Nectin trafficking pathways.} Nectin was internalized via clathrin-mediated endocytosis It was partially retrieved to the PM via Rab11a-mediated recycling or diverted to lysosomal or proteasomal degradation. EE, early endosome; LE, late endosome; Lys, lysosome; Prot, proteasome; RE, recycling endosome. Numbers correspond to the following references: (1) \citep{RN222}, (2) \citep{RN218}, (3) \citep{RN149}, (4) \citep{RN226}, (5) \citep{RN351}, (6) \citep{RN227}, (7) \citep{RN154}.}

\vspace{-0.5cm}

\end{figure*}

The nectin family consists of four single pass junction transmembrane proteins ranging in size from 510 to 549 amino acids \citep{RN208} (Fig. 5). Nectins 1-3 are expressed ubiquitously, whereas nectin-4 is specific to the placenta \citep{RN346}. Nectins form cis-homodimers and either homophilic or heterophilic \textit{trans} dimers via their extracellular domains, which contain three IgG-like loops \citep{RN205}. The nectin cytoplasmic domains, which vary between 41 to 140 amino acids, harbor a carboxy-terminus PDZ-binding motif that conforms to the E/A-X-Y-V (X – any amino acid) consensus sequence. All nectins bind the PDZ domain of the adaptor protein Afadin, which cross links them to f-actin \citep{RN206}. Nectin is the first junction transmembrane protein recruited to the PM during the formation of intercellular junctions \citep{RN153}. Nectin-bound afadin recruits $\alpha$-catenin, followed by cadherin \citep{RN210}. Nectin initiates the formation of tight junctions through the binding of ZO1 to the PDZ domain of afadin \citep{RN211}, which crosslinks nectins by dimerization \citep{RN212}. Afadin’s role in bridging adherens and tight junctions is addressed in section 7 below.

The trafficking pathway of nectin is the least known among junction transmembrane proteins. Nectin was initially identified as a poliovirus receptor \citep{RN213} and means of entry \citep{RN215} into epithelial human cells. Studies on nectin-mediated virus entry indicated that clustered nectins underwent phagocytosis \citep{RN216} or lipid raft-mediated endocytosis \citep{RN215}. Because these mechanisms differ from the clathrin-dependent endocytosis of nectin observed during the remodeling of the  junctions among mouse embryo NIH 3T3 fibroblasts \citep{RN218}, they will not be addressed. Nectin is present in Sertoli cells and was shown to play an important role in spermatogenesis \citep{RN343}. Though it does not function in these cells solely as an intercellular junction protein, it may employ the same endocytic pathway as in epithelial or endothelial cells. The endocytosis of nectin-2 in Sertoli cells was characterized as clathrin-dependent because its degradation was blocked by shRNA-mediated clathrin knockdown \citep{RN225}. Nectin-4 bound in \textit{trans} to nectin-1 on the surface of adjoining A-431 cells underwent transcytosis into the nectin-1-expressing host cell \citep{RN149}. The dominance of nectin-1 in the pulling contest with nectin-4 was attributed to its stronger anchoring to the cytoskeleton relative to nectin-4. The endocytosed nectin-1-nectin-4 dimer apparently escaped degradation as it was not collocated with LAMP-1. Instead, the dimer collocated with vacuolar protein sorting-associated protein (VPS) 35 subunit of the retromer, a trimeric complex that mediates tubulation and sorts cargo proteins for delivery either to the PM or the TGN \citep{RN345}.

Similar to its endocytosis, nectin recycling is poorly known. The recruitment of nectin-1 to the junctions of transformed monkey kidney fibroblasts (COS-7) depended on its binding to MAGUK p55 protein (MPP)-3 \citep{RN222}, a scaffold protein that binds nectin-1 and -3 via its single PDZ domain. The exclusion of nectin-2 indicates that this interaction is highly specific: the last four carboxy-terminus amino-acids of both human and mouse nectin-1 and -3 are identical, but they share only the last two amino-acids with nectin-2. Consequently, MPP3 was not required for cell junction recruitment of nectin-2. In contrast, MPP5, which has a similar domain structure to MPP3, including a single PDZ domain, recruited all three nectins, albeit with lower efficacy for nectin-1 \citep{RN222}. It is unknown how binding to the afadin PDZ domain, the canonical nectin scaffold protein at the cell junctions \citep{RN206}, is reconciled with MPP3 and MPP5 binding to the nectin carboxy-terminus. Studies on Sertoli cells observed colocation of nectin-2 with Rab5a and Rab11a, indicating it reached early endosome and underwent ‘slow’ recycling to the cells’ adhesive junctions \citep{RN154}. 

An additional layer of complexity is added to the recruitment of nectin to cell junctions by the actin \citep{RN348} and afadin-binding \citep{RN227} scaffold protein drebrin. Knockdown of drebrin reduced the abundance of nectin-2 and -3 at the junctions of HUVECs \citep{RN227}. The binding of drebrin to afadin, mediated by their respective poly-proline regions, was required for nectin stabilization at the junctions. This was indicated by the increased abundance of nectin-2 in early endosomes and in lysosomes upon drebrin knockdown. The stabilization of nectin-2 at the cell junctions required the formation of a nectin-afadin-drebrin complex. Afadin is the component of another trimeric complex consisting of PDZ domain-containing protein (PDZD)-11, a small single PDZ domain protein, and pleckstrin homology domain-containing family A member (PLEKHA)-7 which binds afadin and $\delta$-catenin directly \citep{RN351}. The interactions of PDZZ11 with the PDZ-binding motif of nectins 1 and 3 and with PLEKHA7 stabilize adherens junctions by crosslinking nectin to cadherin via PLEKHA7 \citep{RN226}. Similar to drebrin, either PDZD11 or PLEKHA7 knockdown caused partial loss of nectin-1 and -3 from MDCK cell junctions. The loss was prevented by administration of MG132, implying that nectin was degraded by the proteasome, in addition to the lysosome. PLEKHA7 recruited PDZD11 to cell junctions via the binding of its first WW domain to a site located among the first 44 amino-terminal amino-acids of PDZD11. The PDZ domain of PDZD11 bound directly to the carboxy-terminus of nectin-1 and -3 \citep{RN226}, thus stabilizing their attachment to cell junctions.

The studies reviewed above highlight the exceptionally large number of PDZ domain-containing proteins that nectins bind. In addition to afadin, all nectins bind PKC$\alpha$-binding protein (PICK)1 \citep{RN355}, MPDZ, and pals1-associated tight junction protein (PATJ) \citep{RN356}. Nectin-1 and -3 bind also MPP3, MPP5, PDZD11, and PAR3 \citep{RN352}. Though the functional specificity associated with each nectin binding-partner is unclear, their relatively large number and variety suggest that nectin has versatile context-dependent roles in the regulation of intercellular junctions and in other cellular functions. 
\\[\baselineskip]
\begin{center}
\textbf{3.3. FUNCTIONAL CONSEQUENCES OF TIGHT JUNCTION PROTEIN TRAFFICKING}
\end{center}

The constitutive endocytosis and recycling undergone by both E- and VE-cadherin  \citep{RN91,RN110} apparently maintains steady-state levels of the proteins’ abundance at the cell junctions and junction permeability. Multiple studies showed that VEGF, a major destabilizer of endothelial cell junctions, induced endocytosis of VE-cadherin  \citep{RN4,RN6,RN520}. The inflammatory cytokine bradykinin had the same effect  \citep{RN520}. HGF-induced endocytosis of E-cadherin reduced epithelial cell junction integrity \citep{RN88,RN178,RN90}, and, presumably, monolayer permeability.

\begin{center}
\textbf{4. DISCUSSION}
\end{center}

\textbf{4.1.} \textit{Cell condition.} The vast majority of experiments reviewed above were done on cultured cells, a condition that does not reproduce adequately the in vivo state. MDCK cells have been the predominant system of choice, potentially overlooking heterogeneity among epithelial cells. The ‘calcium switch’ one of the frequent treatments for inducing disassembly of epithelial cell junctions, is non-physiological. In vitro experiments on endothelial cells did not account for the hemodynamic effects, which can elevate VE-cadherin phosphorylation in vivo, and, consequently, vein permeability \citep{RN520}. Intercellular junction maturation and integrity were acutely dependent on cell culture conditions \citep{RN167}, and may have been, therefore, inconsistent across studies and experiments.

\textbf{4.2.} \textit{Specificity versus mutuality among endocytic pathways.} Occludin and E-cadherin contain putative sorting motifs in their carboxy-terminus cytoplasmic domains \citep{RN14} that conform to known consensus sequences \citep{RN498}, but their function has not been confirmed. The co-endocytosis of GFP-claudin-3 and endogenous claudin-4 reported by Matsuda et al. \citep{RN54} agrees with the high homology or identity of the putative sorting motif in their 2nd (AF/LGVLL) and 4th (YVGW) transmembrane domain, respectively, proposed by Ivanov et al. Similarly, the sorting motifs in the 4th transmembrane domains of claudins 1, 2, and 7, which were reported to be co-endocytosed by Gehne et al. \citep{RN233} are also either identical (claudins 1 and 2; YLGI) or homologous (claudin-7; FIGW).

A tight versus adherens junction specificity is conferred by the mediation of their recycling by different Rab GTPases. Rab8 GTPases mediate the recycling of adherens junction integral proteins, whereas Rab13 performs the same function in tight junctions \citep{RN102}. Their mutual effector, MICALL2, is shared, however, by both types of cell junction. It is unknown how the junction-type specificity is conferred on Rab8a and Rab13. 

Possibly the most prominent and best understood link between adherens and tight junction proteins is mediated by the adaptor protein afadin. Once bound to nectin, which is the first transmembrane protein recruited to cell junctions, it recruits ZO1 \citep{RN105} via the binding of its proline-rich motifs to the SH3 domain of ZO1 \citep{RN104}. Afadin-bound ZO1 recruits JAM-A \citep{RN106} and claudins, initiating their polymerization into strands \citep{RN65}. Conversely, ZO1/2 facilitate adherens junction assembly \citep{RN499} by participating in the generation of a structurally supportive circumferential f-actin bundle along the cell junctions \citep{RN501}.

Within adherens junctions, the endocytosis of E-cadherin is regulated by nectin. Nectin-bound afadin recruits the GTPase Rap1 \citep{RN500}, inducing the binding of $\delta$-catenin to afadin, and, concomitantly, recruiting E-cadherin. When bound in \textit{trans} to nectin on adjoining cells, nectin blocked E-cadherin endocytosis in its host cell \citep{RN152}. 

\textbf{4.3.} \textit{Fate decision.} Despite having been published more than ten years ago \citep{RN484,RN483}, the insightful discussions of potential fate ‘decision’ mechanisms of integral cell junction proteins are largely relevant today because there has been relatively little progress in understanding these mechanisms. The earliest fate-determining step of cadherin appears to be Src-mediated phosphorylation, followed by ubiquitination, sorting by HRS, and Rab7-dependent trafficking to lysosomes, as described above \citep{RN116}. The endosome outer membrane-binding protein and Rag GTPase GEF LAMTOR1 promoted E-cadherin recycling from early endosomes to the PM, while preventing its trafficking to late endosomes \citep{RN492}. It was not specified, however whether it achieved this effect by activating Rag, or via another mechanism. A different early endosome-associated sorting activity was attributed to the retromer adaptor proteins Snx1, which diverted E-cadherin from degradation to recycling to the PM \citep{RN163}. Other studies observed diversion of E-cadherin from lysosomal degradation to recycling and recruitment to cell junctions following its deubiquitination in the Golgi apparatus by FAM \citep{RN494} or activation of Arf in the TGN by the GEF BIG2 \citep{RN485}, resulting in E-cadherin recyling to the PM \citep{RN373}. 

\begin{figure*}[t]
\center
\includegraphics[scale=0.7]{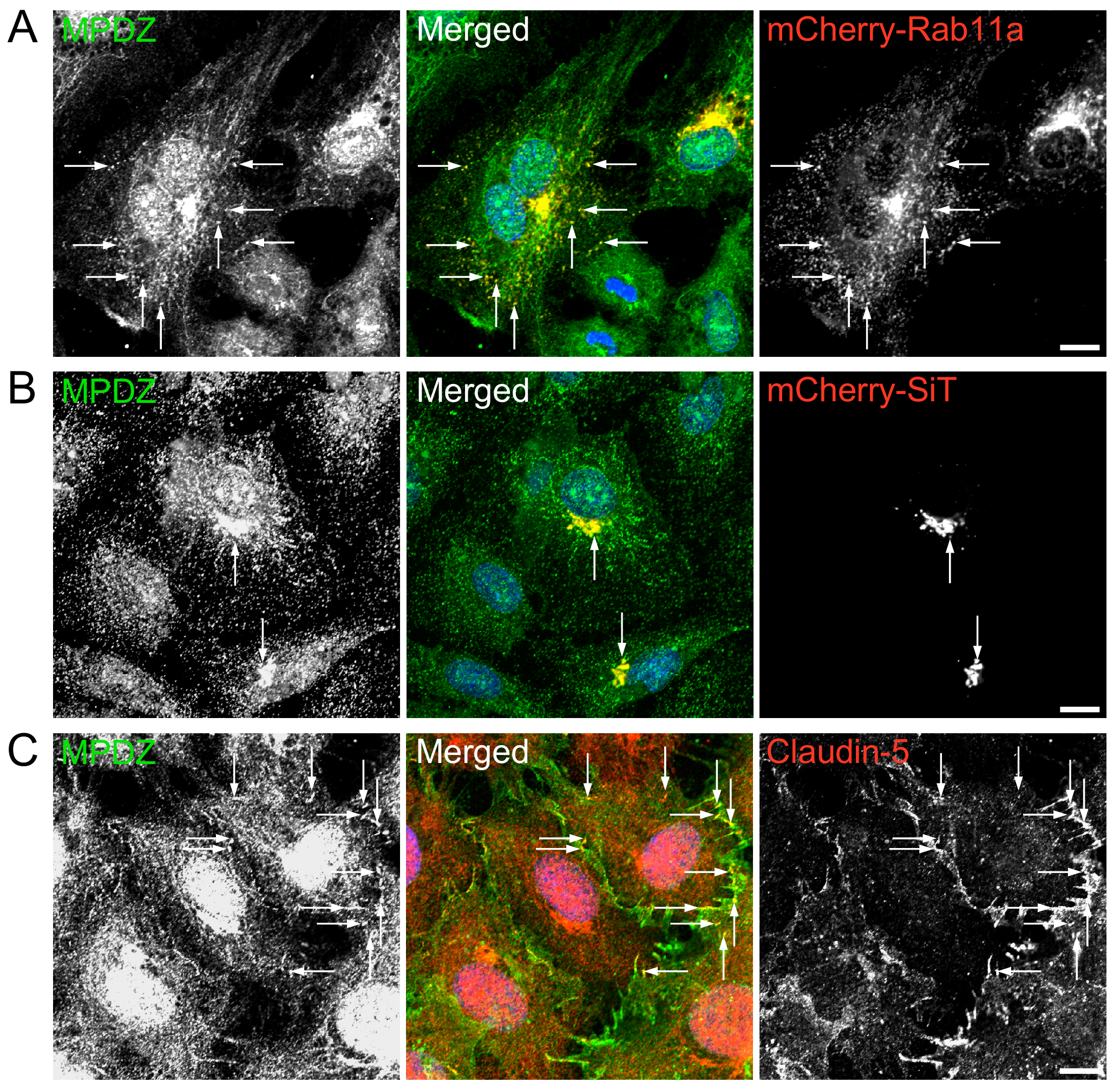}
\vspace{-0.0cm}

\caption{\textbf{MPDZ collocates with vesicular Rab11 and the TGN, and with claudin-5.} Subconfluent human dermal microvascular primary ECs were transduced by lentivirus expressing either Rab11a (A) or sialyltransferase (SiT) (B), a TGN marker \citep{RN504} fused to mCherry, permeabilized by Triton X100, fixed by formaldehyde, and immunolabelled by antibody to MPDZ. Arrows point to cytoplasmic punctae of collocated MPDZ and Rab11 in A, or of collocated MPDZ and TGN marker in panel B. Note that MPDZ and Rab11a collocate also in the TGN. C. ECs permeabilized and fixed as above after 30 min of VEGF treatment were immunolabeled by antibodies to MPDZ and claudin-5. Arrows point to collocated MPDZ and claudin-5 along the cell junctions and in cytoplasmic punctae. Bars, 10 $\mu$m.}

\vspace{-0.5cm}

\end{figure*}

\textbf{4.4.} \textit{Implications of liquid-liquid phase separation.} Recent studies demonstrated that the scaffold protein ZO1 drives the generation of tight junctions while forming cytoplasmic condensates in live MDCK cells \citep{RN381}. The formation of such condensates by ZO1 and potentially by other junction-associated scaffold proteins, e.g., MPDZ and PATJ \citep{RN481}, is a novel aspect of intercellular junction dynamics that provides new insight into their assembly mechanisms. The PM, ER, and endosomes provide platforms for the formation of liquid-liquid phase-separated condensates \citep{RN469}. The emerging mechanism of tight junction initiation consists of ZO1 transition into PM-bound condensates, the partitioning of claudin, occludin, and of the scaffold proteins afadin and cinglulin into the condensates, and the triggering of claudin polymerization \citep{RN381}. Do claudin and occludin, which translocate from the TGN to the PM by membrane trafficking, dissociate from the vesicles that carry them and partition into ZO1 condensates, or do they partition into the ZO1 condensate while remaining inserted into the vesicles’ membranes until the vesicles fuse with the PM?

\textbf{4.5.} \textit{Functional specificity of scaffold proteins.} As elaborated above, the ZO protein family plays a major role in the recruitment of tight junction proteins as well as in the structural support of tight junctions, whereas afadin plays a similar role in adherens junction assembly. ZO1/2 bind numerous claudins via their PDZ domains. Afadin binds nectins in the same manner. These, however, are not the only binding partners of the PDZ-binding motifs of claudins, JAMs, and nectins. The homologous large scaffold proteins PATJ and MPDZ bind essentially the same junction transmembrane proteins and have been localized to the cell junctions, similar to ZO1/2 and afadin \citep{RN502,RN503}. How then are the interactions of claudins and nectins with ZO1/2 and afadin, respectively, reconciled with their interactions with PATJ/MPDZ? What are the specific functions of PATJ/MPDZ? Are those functions linked to intercellular junction homeostasis similar to ZO1/2 and afadin? While these questions cannot be answered in full at this time, preliminary data from the author’s lab show that MPDZ collocates with Rab11a in cytoplasmic punctae (Fig. 6A) and is present in the TGN of subconfluent human dermal primary microvascular ECs (Fig. 6B), suggesting that MPDZ undergoes membrane trafficking from the TGN to cell junctions. In parallel, MPDZ collocates with claudin-5, a prominent claudin in ECs \citep{RN508}, along EC junctions and in cytoplasmic punctae (Fig. 6C), suggesting that MPDZ, and possibly PATJ, are involved in junction protein dynamics rather than in their stabilization at the cell junctions. 
\\[\baselineskip]
\textbf{Acknowledgement:} I attempted to review all the studies I considered relevant to the review’s subject. Studies that were superseded by more recent ones were not included. Readers are encouraged to inform me of missed studies. I will consider all suggestions and include them if they add new information.
\\[\baselineskip]
\textbf{Conflict of interest:} None.

\typeout{}
\bibliographystyle{abbrvnat}
\bibliography{References}

\end{document}